\documentclass[sigconf]{acmart}

\AtBeginDocument{%
  }

\setcopyright{acmlicensed}
\copyrightyear{2018}
\acmYear{2018}
\acmDOI{XXXXXXX.XXXXXXX}

\acmConference[Conference acronym 'XX]{Make sure to enter the correct
  conference title from your rights confirmation emai}{June 03--05,
  2018}{Woodstock, NY}
\acmISBN{978-1-4503-XXXX-X/18/06}

\usepackage{xcolor}
\usepackage{skeldoc}
\usepackage{svg}
\usepackage{subcaption}
\usepackage{tabularx}
\usepackage{svg}
\usepackage{booktabs} 
\usepackage{balance} 
\usepackage{soul} 
\copyrightyear{2024}
\acmYear{2024}
\setcopyright{acmlicensed}\acmConference[KDD '24]{Proceedings of the 30th ACM SIGKDD Conference on Knowledge Discovery and Data Mining}{August 25--29, 2024}{Barcelona, Spain}
\acmBooktitle{Proceedings of the 30th ACM SIGKDD Conference on Knowledge Discovery and Data Mining (KDD '24), August 25--29, 2024, Barcelona, Spain}
\acmDOI{10.1145/3637528.3671532}
\acmISBN{979-8-4007-0490-1/24/08}
\begin{document}

\title{Television Discourse Decoded: Comprehensive Multimodal Analytics at Scale}

\author{Anmol Agarwal}
\authornote{Equal contribution.}
\affiliation{%
  \institution{International Institute of Information Technology}
  \city{Hyderabad}
  \country{India}
}
\email{anmol.agarwal@students.iiit.ac.in}

\author{Pratyush Priyadarshi}
\authornotemark[1]
\affiliation{%
  \institution{International Institute of Information Technology}
  \city{Hyderabad}
  \country{India}
}
\email{pratyush.priyadarshi@students.iiit.ac.in}

\author{Shiven Sinha}
\affiliation{%
  \institution{International Institute of Information Technology}
  \city{Hyderabad}
  \country{India}
}
\email{shiven.sinha@research.iiit.ac.in}

\author{Shrey Gupta}
\affiliation{%
  \institution{International Institute of Information Technology}
  \city{Hyderabad}
  \country{India}
}
\email{shrey.gupta@students.iiit.ac.in}

\author{Hitkul Jangra}
\affiliation{%
  \institution{Indraprastha Institute of Information Technology}
  \city{Delhi}
  \country{India}
}
\email{hitkuli@iiitd.ac.in}

\author{Ponnurangam Kumaraguru}
\affiliation{%
  \institution{International Institute of Information Technology}
  \city{Hyderabad}
  \country{India}
}
\email{pk.guru@iiit.ac.in}

\author{Kiran Garimella}
\authornote{Corresponding author}
\affiliation{%
  \institution{Rutgers University}
  \city{New Brunswick}
  \country{USA}
}
\email{kiran.garimella@rutgers.edu}

\renewcommand{\shortauthors}{Anmol Agarwal et al.}
\newcommand{\answerYes}[1]{\textcolor{blue}{#1}} 
\newcommand{\answerNo}[1]{\textcolor{teal}{#1}} 
\newcommand{\answerNA}[1]{\textcolor{gray}{#1}} 
\newcommand{\answerTODO}[1]{\textcolor{red}{#1}} 
\definecolor{lightred}{rgb}{1,0.6,0.6}
\definecolor{darkgreen}{rgb}{0,0.5,0}
\definecolor{lightgreen}{rgb}{0,0.8,0.1}

\newcommand{\striketext}[1]{%
    \textcolor{lightred}{\st{#1}}%
}
\newcommand{\sensitive}[1]{%
    \textcolor{blue}{#1}%
}
\newcommand{\gitdiff}[2]{%
  \striketext{}%
  {#2}%
}
\newcommand{\gitdifftwo}[2]{%
  \striketext{#1}%
  \textcolor{lightgreen}{#2}%
}

\newcommand{\channel}{\textsc{the Channel}}
\newcommand{\anchor}{\textsc{the Anchor}}
\newcommand{\tvshow}{\textsc{the Show }}
\newcommand{\todoadd}{\textbf{\textsc{todo add}}}
\newcommand{\supplement}{supplemental pdf}
\begin{abstract}

In this paper, we tackle the complex task of analyzing televised debates, with a focus on a prime time news debate show from India. Previous methods, which often relied solely on text, fall short in capturing the multimodal essence of these debates~\cite{proksch2019testing}. To address this gap, we introduce a comprehensive automated toolkit that employs advanced computer vision and speech-to-text techniques for large-scale multimedia analysis.
Utilizing state-of-the-art computer vision algorithms and speech-to-text methods, we transcribe, diarize, and analyze thousands of YouTube videos of a prime-time television debate show in India. These debates are a central part of Indian media but have been criticized for compromised journalistic integrity and excessive dramatization~\cite{oxIndianNews}.
Our toolkit provides concrete metrics to assess bias and incivility, capturing a comprehensive multimedia perspective that includes text, audio utterances, and video frames. Our findings reveal significant biases in topic selection and panelist representation, along with alarming levels of incivility. This work offers a scalable, automated approach for future research in multimedia analysis, with profound implications for the quality of public discourse and democratic debate. To catalyze further research in this area, we also release the code, dataset collected and supplemental pdf\footnote{\href{https://github.com/anmolagarwal999/television-discourse-decoded}{https://github.com/anmolagarwal999/television-discourse-decoded}}.

\end{abstract}



\maketitle

\section{Introduction}
Television debates are a cornerstone of public discourse, serving as platforms for the exchange of ideas and viewpoints. In India, prime-time debates are viewed by millions and have a substantial impact on shaping public opinion~\cite{bhat2023expanding}. However, these debates have recently undergone scrutiny for compromised journalistic integrity and increasing incivility~\cite{mishra2018broadcast}. Understanding the nuances at scale in these debates is important, yet a formidable task due to the multimedia nature of the content, which blends text, audio, \& video.

Automated methods to analyze such content have largely been absent or inadequate, often focusing only on textual aspects~\cite{proksch2019testing}. These naive approaches are insufficient for two reasons: the sheer scale of televised debates available for analysis, and the intricate multimedia elements that must be considered to provide a complete picture. Previous attempts at solving this problem either employ text-based analytics that miss out on contextual cues or rely on small-scale, manual coding that lacks scalability \cite{joo2019automated,konstantinovskiy2021toward}.

One of the most intriguing yet challenging aspects of analyzing news debates lies in their multimodal nature, which combines text, audio, and visual elements. Each of these modalities carries crucial information that contributes to the complete understanding of a debate. While text may convey the spoken content, it misses out on the tone, pitch, and interruptions that audio captures. Similarly, video offers visual cues like facial expressions and body language that are lost in a purely textual analysis. Thus, a comprehensive analysis mandates a multifaceted approach that considers all these elements in unison.

Scale further complicates this endeavor. The vast number of televised debates—spanning thousands of episodes and millions of minutes of footage—requires a computational approach capable of scaling without loss of accuracy. Moreover, the temporal dynamics intrinsic to debates, such as topic changes and emotional fluctuations, add another layer of complexity. Capturing these dynamics over time demands sophisticated algorithms that can adapt to fast-changing contexts within a debate.

Beyond the technical aspects, subjective elements like bias and incivility pose their own challenges. Creating universally applicable metrics for these elements is particularly difficult, as perceptions of bias can differ based on individual viewpoints. Similarly, cultural and linguistic nuances like local idioms or specific styles of argumentation, especially pertinent in the Indian context, require additional considerations for accurate analysis. The presence of speech overlaps and interruptions further muddies the waters. These not only challenge the speech-to-text conversion process but also have implications for downstream analytics, potentially affecting the quality of the transcriptions and, consequently, the entire analysis. In cases where real-time analysis is required, these complexities amplify, adding an additional computational burden.

In light of these challenges, this paper introduces a novel automated toolkit designed for large-scale multimedia analysis. Our approach leverages state-of-the-art advances in computer vision algorithms and speech-to-text methods to transcribe, diarize, and analyze thousands of televised debates hosted on YouTube. We collect data spanning over 6 years from one of India's most popular prime time news debate shows (henceforth referred as \tvshow) which airs on one of India's most watched English news channel (henceforth referred as \channel).
The program is known for its emphasis on nationalistic discourse, its fervent critique of political adversaries, and its often intense discussions involving minority groups. It is widely perceived that the channel exhibits a preference for the incumbent government; however, this perspective has not been rigorously substantiated through quantitative research methods.

To fill this gap, we offer concrete metrics to evaluate bias in discussion topics and measure levels of incivility. Furthermore, our toolkit amalgamates textual transcriptions with video frames and audio utterances, thus capturing a comprehensive multimedia perspective. This offers a much-needed foundation for future research, making it possible to conduct studies that are both wide-ranging and deep in their analytical scope.

Our work is situated within the broader, ongoing debate about the quality of television debates in India, which have recently come under criticism for a rise in sensationalism, dramatization, and incivility. We capture these elements in our analysis to provide a comprehensive understanding of the current state of televised debates in the country.

Our analysis reveals a degree of bias in the debate show, characterized by support for the Ruling Party and a tendency to take a discrediting stance towards opposition parties and journalists. Our study also points out a notable imbalance between male and female panelists, leading to an uneven representation of social issues. It's especially concerning to see the high levels of rudeness measured by our pipeline: on average, shouting happens in about 9\% of the duration of the videos. These findings have profound implications. The pronounced bias and a lack of dignified discourse cast doubt on the show's role as a fair platform for different opinions.
This calls into question the show's role in fostering constructive public debate; instead, it appears to prioritize sensationalism, potentially at the cost of nuanced discussion and mutual understanding.

We make both our data analysis pipeline and the collected data publicly available. This is expected to catalyze further research in automated video analysis, extending its applicability beyond the Indian context. 
By doing so, we aim to unlock the untapped potential of YouTube as a tractable resource for large-scale studies.
\section{Background and Related Work}

\subsection{Bias and Incivility in Indian media}

India, the world's largest democracy, has recently experienced a decline in press freedom, currently ranking 161 out of 180 countries as per Reporters Without Borders~\cite{rsfIndia}. This decline has been partly attributed to the acquisition of media outlets by influential figures who maintain close ties with political leaders. Such ownership structures have led to seemingly evident biases in media reporting, with a majority of TV channels noticeably supporting the political party in power. Given the critical role of media in a democratic setup, it becomes imperative to analyze and quantify this bias, a task that some previous work has approached qualitatively.

Since \channel's inception, it has been the most-watched English news channel in India, commanding an average viewership of 40\% \cite{broadcast_audience_research_council_data_2022}. Known for its sensationalist approach to news reporting, \channel~ has often been criticized for displaying a pro-Hindu, pro-nationalist, and pro-government bias \cite{subhajit2021media}.
One of the channel's popular prime time debate show, henceforth, referred as \tvshow epitomizes this tendency. The show attracts over five million daily viewers and is characterized by its nationalistic tone. It often targets those who appear to oppose the government's viewpoint.
\\
Despite its status as the most-watched news TV show in India, the program has moved away from the traditional format of a balanced news debate. Instead, it now often features a heightened level of dramatization, impassioned language, and overlapping dialogue~\cite{oxIndianNews}. This sensational approach appears to resonate with viewers ~\cite{scrollWatchArnab}.

While there is a substantial body of qualitative work addressing bias, factual inaccuracies, and the dramatization of news in Indian media~\cite{drabu2018muslim,mishra2018broadcast,bhat2023expanding}, our research contributes by offering quantitative evidence. Notably, some channels, including \channel, have even acknowledged their tendencies to sensationalize news. Our study enriches this dialogue by supplying empirical data on the nature and framing of the content presented in such debate shows.

\subsection{Analysis of TV News and Media}
\begin{figure*}[htbp]
    \centering
    \includegraphics[width=\linewidth]{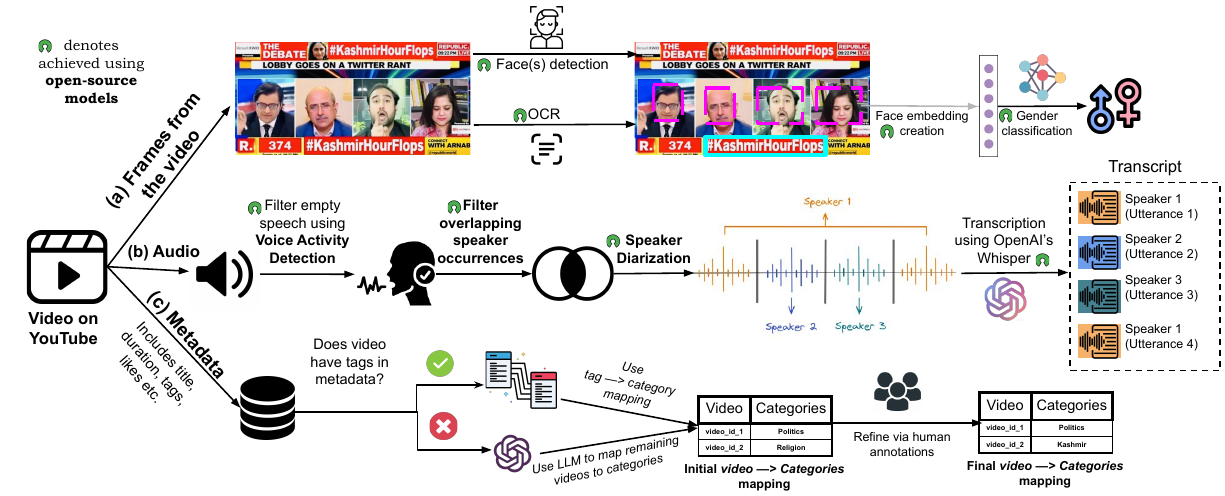}
    \caption{Pipeline overview: \textbf{Branch (a)} details the process for identifying gender from facial data in videos and extracting hashtags from debate screens; \textbf{Branch (b)} outlines the audio cleaning and speaker diarization procedures, followed by transcription of utterances into text; \textbf{Branch (c)} illustrates the semi-automated annotation system that leverages YouTube metadata \& LLMs to streamline the categorization of videos into categories, thereby reducing human annotation workload.}
    \label{fig:pipeline_fig}
\end{figure*}
In the realm of analysis of TV news and media, multiple avenues of research have emerged that address the intricate problem of media bias, the influence of media on public perception, and the role of technological platforms in shaping or amplifying these biases. One stream of work delves into detecting subtle biases in online news by examining `gatekeeping,' coverage, and statement bias, using unsupervised methods on a geographically diverse set of news sources~\cite{saez2013social}. This line of research intersects with another that undertakes a comparative framing analysis of terrorism coverage in US and UK newspapers, revealing differing national focuses, either militaristic or diplomatic, that guide news stories' framing~\cite{papacharissi2008news}.

While these studies examine traditional media forms, a more recent shift towards social media as a news outlet is apparent in the research literature. For example, some researchers employ scalable methodologies that leverage social media's advertiser interfaces to infer the ideological slant of thousands of news outlets. This method provides granularity, capturing demographic biases that go beyond political leanings, and results in deployable systems for transparency~\cite{ribeiro2018media}. This complements work on newspaper endorsements' influence on voting behavior, highlighting source credibility as a key factor in endorsement effectiveness.

Interestingly, research has also been conducted in the Indian context, where media bias in policy coverage has been systematically quantified. This work reveals biases in topic selection and representation of different social classes and political parties. Notably, social media platforms seem to echo rather than mitigate these biases, an insight that aligns with the earlier observations on the role of social media in amplifying traditional media biases~\cite{sen2022analysis}. Collectively, these studies illuminate the evolving landscape of news and media analysis, showcasing the need for comprehensive, multifaceted approaches. They underline the significance of understanding both the subtleties in traditional media framing and the influential role of social media platforms.

\subsection{Multimodal Analysis Tools}
Video analysis has become an increasingly significant area of research, particularly as social media platforms transition towards video-centric content. The rise of short video services like TikTok underscores the growing importance of video in the digital age. Advances in computer vision technology have reached a stage where real-world applications are not just feasible but increasingly sophisticated. Problems such as video summarization and key frame extraction have been addressed, offering novel solutions and methodologies~\cite{DBLP:journals/air/SainiKKSN23, liu2003novel}.

Earlier works ~\cite{beeferman2019radiotalk} faced challenges in transcribing large volumes of audio data—284,000 hours of radio—due to the limitations in transcription models at the time. The current models for transcription have improved considerably showcasing a rapid evolution of the field.
Videos present a complex interplay of multiple modalities, including visuals, text, and audio. While each can be analyzed independently, their true power lies in how they interact. \citet{renoust2016visual} explored this by using deep neural networks for face detection and text counting metrics to measure politicians' screen time. Their work demonstrated the capability of AI techniques in analyzing large video datasets, offering insights into complex social dynamics.
The GDELT Project~\cite{gdeltprojectGDELTSummary} provides web-based interfaces for analyzing caption text and other on-screen elements but lacks in-depth labelling related to voice tone or content being discussed. 

Our work fills these gaps by analyzing a comparable dataset of videos and enriches it by labelling content related to what is spoken, who is on-screen, and the tone of voice used.
Overall, our research builds on recent advancements in various domains of AI. We leverage state-of-the-art models in image processing for tasks such as face and gender recognition, utilize speech processing algorithms to identify instances of shouting, and employ speech-to-text models to capture the spoken content. We aim to provide a holistic, multi-modal analysis that can serve as a robust foundation for future studies in video analytics.
\section{Data Collection \& Processing}
Our primary dataset comprises 2,087 hours of debate footage from 3,000 videos. Initially, we used the YouTube Data API\footnote{\url{https://developers.google.com/youtube/v3/docs/playlistItems/list}} to extract metadata from the official playlist of the ~\tvshow~as of December 2022. This provided us with 3,151 unique videos dating back to May 2017.
Out of these, we filtered out $67$ videos because they were too short/long (i.e. their duration was less than $10$ minutes or exceeded $4$ hours) and filtered out an additional $84$ videos because the annotators couldn't agree on their categories. We were finally left with 3,000 videos corresponding to over 2,087 hours of video content. The metadata fetched using the YouTube Data API for each video contains the title, URL, description, and a list of tags chosen by the channel\footnote{Details: \url{https://developers.google.com/youtube/v3/docs/videos\#snippet.tags[]}} associated with the video.

\subsection{Categorizing the Videos}
To categorize the 3,000 videos in our dataset, we manually created categories. Initially, using a framework from a prior study we adopted 18 categories ~\citep{caravan_article}. 
Each coder independently assessed a subset of videos, relying on metadata such as titles, descriptions, hashtags, and tags for initial categorization. If a video did not fit into the existing categories, a new category was proposed and discussed among coders for potential inclusion. This iterative process continued until a consensus was reached on the categories.

Recognizing that a video could span multiple topics, we implemented a two-tiered coding system comprising major and minor categories. Each video was assigned to one major category (e.g., sports, religion, international affairs) while potentially belonging to multiple minor ones, allowing for emergent sub-themes (e.g., `Russia-Ukraine crisis', `SSR case').
To automate video categorization, we used the "tags" present in the YouTube metadata. The tags were then mapped to categories. For example, a video with tag "budget 2019" was labelled Economy as the major category.
However, after this initial categorization, we were left with 830 videos that could not be mapped to a category due to the absence of tags or the presence of generic tags. We then used GPT-4 to map these remaining videos to the categories based on the video's title. 

The final step involved human refinement to correct any potential errors from the automated labelling. Two annotators independently examined and refined the labels, and their agreement was measured using the Fleiss kappa statistic, which was computed to be 0.933, indicating excellent agreement. 
By incorporating LLMs and tags-metadata we reduced the number of categories that can be mapped to a video to a smaller subset thereby significantly reducing the time taken in the human annotation. This hybrid approach of automated and human annotation in our pipeline allowed for an efficient and comprehensive categorization of the videos. \\
In a minority of the cases with disagreements (110 cases), both the annotators discussed among themselves and resolved most of the disagreements. There was no clear agreement on 84 videos which were removed from further analysis, leaving us with $3,000$ videos. A complete breakdown of major and minor categories is available in Appendix \ref{sec:additionaldetails}. The majority of the videos fall into five dominant categories: Politics, Religion, COVID Lockdowns, International Affairs, and Crime \& Justice, collectively accounting for 66\% of the total dataset. A mapping from these categories to their respective tags and examples of the annotation process can be found in the \supplement. Our semi-automated pipeline has been illustrated in branch (c) of Figure \ref{fig:pipeline_fig}. 

\subsection{Transcription and Speaker Diarization} 
\label{sec:transcription_diarization}
To analyze the content of the debates, it was essential to determine both what was said and who said it. Audio transcription converts speech in an audio file into written text, but debates involve multiple speakers in multi-turn interactions. Therefore, before transcription, we performed speaker diarization—a process that partitions an audio stream into segments and attributes them to specific speakers~\cite{park2022review}. This allowed us to transcribe individual speaker segments, resulting in a conversation-format transcription for each video.

We executed two key pre-processing steps to enhance the quality of the diarization results. First, we removed segments devoid of speech, such as interstitials and speaker transitions, using the \textbf{Voice Activity Detection} feature from the Pyannote toolkit~\cite{bredin2020pyannote}. This removal improved subsequent diarization accuracy. Second, we filtered out overlapping speech segments to avoid performance degradation in speaker clustering during diarization, accomplished using the same Pyannote model~\cite{bredin2021end}.

After these pre-processing steps, we employed the Pyannote \textbf{diarization} module to partition the audio into homogeneous segments, each assigned to a specific speaker~\cite{bredin2020pyannote, bredin2021end}. For transcription, we leveraged OpenAI's Whisper speech-to-text model~\cite{radford2023robust}, notable for its robust performance on diverse accents and technical language. Whisper has demonstrated near-human-level accuracy in challenging noisy settings~\cite{li2023efficient}. Combining Whisper's transcription capabilities with Pyannote's audio segmentation and speaker diarization enabled us to transcribe and accurately attribute speech (and the corresponding transcribed text) to individual speakers.

Our qualitative analysis revealed certain limitations in the Pyannote model's \textbf{overlap detection}. Specifically, the model only considered speech overlapping if all audio segments were incoherent. If one speaker's voice dominated others, the model did not recognize the speech as overlapping. This issue resulted in scenarios where multiple speakers are active, but not identified as overlapping. Additionally, the transcription quality for overlapped speech was suboptimal, likely because Whisper's training data primarily focuses on transcribing a single speaker while treating other voices as background noise.\footnote{\url{https://github.com/openai/whisper/discussions/434\#discussioncomment-4141250}}
Due to these overlap detection limitations, we encountered `spurious speakers'---artifacts that appeared to be individual speakers but were actually combinations of multiple voices. Such spurious speakers also emerged when the debate anchor played relevant footage with accompanying audio, complicating the speaker diarization process.
Nevertheless, this might impact a small fraction of our video content and manual evaluations on a subset of videos showed that the overall quality of the transcripts was exceptional. The entire transcription pipeline is outlined in \textit{Branch (b)} of Figure \ref{fig:pipeline_fig}.

\subsection{Face and Gender Detection}
Gender identification from video frames, as shown in \textit{Branch (a)} of Figure \ref{fig:pipeline_fig}, entailed extracting and analyzing facial data. For facial recognition in our study, we employed the DeepFace library~\cite{serengil2020lightface}, specifically utilizing the RetinaFace detector coupled with the VGG-Face model~\cite{serengil2021lightface}. From a given video, we sampled one frame every 3 seconds and extracted all faces from it. One challenge we encountered was the presence of spurious faces, in advertisements or images unrelated to the debate. To address this, we implemented a filtering mechanism based on the size of the face in the frame and the confidence scores provided by the model. It's important to acknowledge that our study operates within the limitation of recognizing gender in binary terms, although we recognize that gender is not a binary construct. In a small-scale experiment to validate the performance of this model, we annotated all the faces on 2,500 randomly sampled frames across our dataset and found the classifier to have a precision of 0.91 and a recall of 0.994 for males, and a precision of 0.975 and a recall of 0.81 for females.


\subsection{Extracting Panelist Names from Transcripts}
\label{sec:panelist_names}
To study the individuals appearing in the debates, we extracted the names of panelists from the transcripts. Traditional approaches like Named-Entity Recognition (NER) on the transcripts did not perform well for three main reasons: (i) NER captured names of people mentioned in the debate but not actually panelists, (ii) multiple variations were used to refer to the same person (e.g., [General GD Bakshi, General Bakshi, Major General GD Bakshi]), and (iii) transcription errors led to inconsistent spellings of the same name (e.g., Atiqur Rahman, Atiq-ur-Rehman Sahab, Atiku Rehman). To address these issues, we adopted Meta's open-sourced LLaMA-2 13B model for this task~\cite{llama_2_touvron2023llama}.

When the transcript of an entire video exceeded the model's context length, we chunked the transcript into parts and took the union of names extracted from each chunk to identify potential panelists for the video. The prompt used for name extraction can be found in the \supplement. The names returned by this approach were not completely clean, so we performed fuzzy matching and clustered similar names using a combination of Partial Token Sort Ratio and metaphone-based matching.

Using these techniques, we curated a list of 265 panelists, covering 91.7\% of the videos and 50\% of all appearances. We focused on frequently invited guests rather than full coverage due to the long tail distribution of debate participants. To validate our pipeline, one author manually identified panelists in 50 videos and compared them to our pipeline's results, achieving a precision of 0.901 and recall of 0.730.

Next, we identified and coded the occupation of the panelists into categories such as TV-related, academics, activist, advocate, analyst, author, civil servant, consultant, doctor, film-related, journalist, politician, religious leader, social leader, and spokesperson. We also coded their affiliations (e.g., political party support).

From the initial set of 285 people identified, 20 were removed as false positives. We only marked individuals who were part of some organization (e.g., Bombay High Court, Samajwadi Party, DMK, BJP, All India Trinamool Congress, \channel, Congress) and marked 'None' for others.
\section{What is discussed in the debates?}

\subsection{Bias in Transcripts}
\begin{table*}[h]
    \centering
    \small
    \caption{Words found to be important in the context in sentences involving the Ruling Party and the Opposition. (* indicates that the word was not present in BERT vocabulary and the score is indicative of the word's subtokens. Eg: raf$\rightarrow$rafale, par$\rightarrow$parivar)}
    \label{tab:bjp_ig}
    \setlength{\tabcolsep}{3pt} 
    \begin{tabular}{|l|l|l|l|l|l|l|}
    \hline
    \multicolumn{7}{|c|}{\textbf{Ruling Party related words}}\\
    \hline
    wave (\textit{0.645}) & hate (\textit{0.635}) & trump (\textit{0.603}) & hatred (\textit{0.595}) & bengal (\textit{0.573}) & factor (\textit{0.517}) & ji (\textit{0.501}) \\ \hline
    pm (\textit{0.483}) & model (\textit{0.443}) & cabinet (\textit{0.4}) & voted (\textit{0.397}) & defeat (\textit{0.375}) & riot (\textit{0.362}) & vote (\textit{0.354}) \\ \hline
    2019 (\textit{0.354}) & uttar (\textit{0.321}) & kashmir (\textit{0.308}) & rallies (\textit{0.306}) & responsible (\textit{0.269}) & victory (\textit{0.264}) & pakistan (\textit{0.262}) \\ \hline
    secular (\textit{0.259}) & development (\textit{0.252}) & power (\textit{0.248}) & democracy (\textit{0.247}) & policy (\textit{0.232}) & poll (\textit{0.231}) & elected (\textit{0.198}) \\ \hline
    economy (\textit{0.197}) & farmers (\textit{0.167}) & global (\textit{0.164}) & campaign (\textit{0.156}) & 2014 (\textit{0.154}) & security (\textit{0.142}) & credit (\textit{0.133}) \\
    \hline
    \multicolumn{7}{|c|}{\textbf{Opposition related words}}\\
    \hline
    indira (\textit{0.772}) & baba (\textit{0.473}) & mother (\textit{0.444}) & dynasty (\textit{0.442}) & rafale \textbf{*} (\textit{0.362}) & apologize (\textit{0.348}) & vatican (\textit{0.344}) \\ \hline
    parivar \textbf{*} (\textit{0.327}) & silent (\textit{0.275}) & victim (\textit{0.272}) & questioning (\textit{0.268}) & lie (\textit{0.262}) & age (\textit{0.26}) & italian (\textit{0.257}) \\ \hline
    courage (\textit{0.256}) & personal (\textit{0.233}) & exposed (\textit{0.231}) & silence (\textit{0.23}) & concerned (\textit{0.22}) & lobby (\textit{0.209}) & son (\textit{0.207}) \\ \hline
    shame (\textit{0.174}) & fake (\textit{0.169}) & brother (\textit{0.168}) & hindus (\textit{0.165}) & secret (\textit{0.161}) & sorry (\textit{0.147}) & evidence (\textit{0.122}) \\ \hline
    president (\textit{0.122}) & investigation (\textit{0.121}) & corruption (\textit{0.116}) & communal (\textit{0.101}) & chinese (\textit{0.092}) & xi-jinping \textbf{*} (\textit{0.088}) & failed (\textit{0.087}) \\ \hline
    
    \end{tabular}
    
\end{table*}
Existing literature ~\cite{caravan_article,subhajit2021media} supports the notion that the show exhibits a pro-government stance. Our categorization, summarized in Appendix \ref{sec:additionaldetails}, aligns with this perspective, revealing a significant 3-to-1 ratio in favor of narratives that support the ruling party. However, unlike previous works, this paper zeroes in further on the \textit{content} of the show to showcase a political tilt, if any. To achieve this, we work with the transcripts and adopt a methodology akin to those in~\cite{Ding,khuda_Palakodety2020MiningIF}, utilizing language models to identify potentially biased attributive/contextual tokens.

Specifically, we train a classifier to determine if a sentence in the transcript pertains to the ruling party or the opposition. This classifier is based on a fine-tuned BERT-Base-Uncased model ~\cite{DBLP:journals/corr/abs-1810-04805}, equipped with a classification head.

For classifier training, we select sentences from the transcripts that explicitly reference the ruling party or the opposition, using specific keywords such as names of parties or leaders (Appendix \ref{sec:additionaldetails}). We exclude sentences that mention both to prevent ambiguity. To ensure the model focuses on the context rather than the keywords, we mask the specific keywords, replacing person names with \textless{}PER\textgreater{} and party names with \textless{}PARTY\textgreater{}.

Given BERT's shortcomings in handling negations~\cite{khandelwal-sawant-2020-negbert}, we exclude sentences containing negation keywords such as \textit{not},\textit{won't} etc. Our final dataset comprises $16,444$ sentences about the Opposition and $14,865$ about the Ruling Party, divided into 80\% training, 10\% validation, and 10\% test sets. The model is fine-tuned for $30$ epochs with a batch size of $32$, using the AdamW optimizer at a 2e$^{-5}$ learning rate.

To make the model's decision-making process more interpretable, we use \textit{integrated gradients} \cite{ig_paper}, a technique that effectively determines the influence of individual tokens on the model's predictions. This approach helps us pinpoint the tokens that significantly sway the model's judgment in classifying sentences as pertaining to the Ruling Party or the Opposition, in line with Ding et al.~\cite{Ding}.

Our classifier achieved an accuracy of $85.72\%$. For a nuanced understanding, we sorted the words in each category by their average attribution scores across all sentences. After excluding stopwords, infrequently occurring words (less than 50 times), and generic terms to minimize noise, a qualitative analysis of these highly-attributable tokens reveals a distinct bias against the Opposition, while favouring the Ruling Party. The complete list can be found in Table \ref{tab:bjp_ig}. Below, we provide examples to illustrate this qualitatively:\\

\noindent \textbf{Ruling Party related tokens:}
\\(i) \textit{Election-centric Narratives}: Tokens like `vote' `victory' and `power' suggest a focus on the electoral successes of the ruling party.
\\(ii) \textit{Veneration of Leadership}: Terms like `Modi wave,' `Modi factor,' and respectful suffixes like `ji' (as in `Modiji') paint a picture of reverence around the party leadership. The term 'development' often co-occurs, framing the ruling party as a catalyst for progress.
\\(iii) \textit{Defensive and Counter-Narratives}: Surprisingly, words like `hatred' appear in the context of disputing the notion that animosity towards ruling party is justified. Other tokens like 'Trump' and 'Pakistan' indicate international validation or emphasize a tough stance on national security.

\noindent \textbf{Opposition related tokens:} \\(i) \textit{Dynastic Politics}: Usage of words like `dynasty,' and familial references like `mother-son-sister' aim to cast the main Opposition party in a light suggestive of nepotism.
\\(ii) \textit{Name-Calling and Stereotypes}: Phrases like `Rahul Baba,' `Vadra Congress,' and references to `lobby' paint the main Opposition party with connotations of naivety \& questionable ethics, or disloyalty.
\\(iii) \textit{Allegations and Scandals}: Terms like `Rafale,' `China,' and `Jinping' are mentioned in contexts that suggest improper or unpatriotic conduct by the main Opposition party. Words like `fake,' `shame,' and `lie' reinforce a narrative of dishonesty and ineptitude. \\

We also find similar bias in hashtags used for the show.
To fetch the hashtags displayed on the screen, we sampled a frame every 30 seconds and extracted text using EasyOCR~\cite{githubGitHubJaidedAIEasyOCR}.
The text corresponding to the hashtags was extracted using a regular expression.
We see a clear pattern in how the hashtags are chosen: while criticisms of the ruling party tend to be issue-specific and nuanced, criticisms of the Opposition are likely to be sweeping and derogatory, contributing to a narrative that could potentially influence public perception.
\begin{figure}
    \centering
    \includegraphics[width=1\linewidth, clip=true, trim=0 35 0 100]{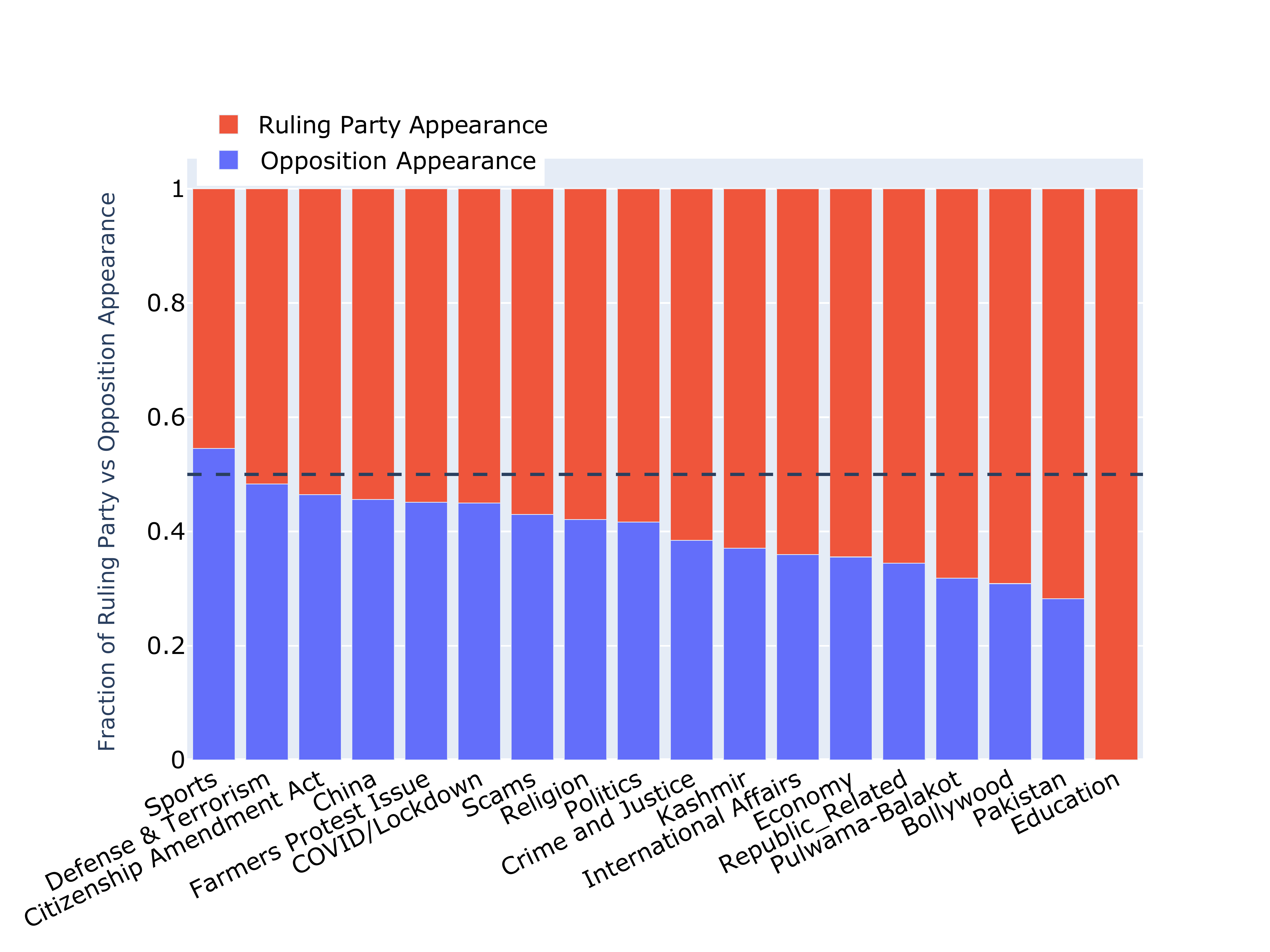}
    \caption{Fraction of panelists invited from the ruling party vs. the opposition. Pro-ruling-party panelists appear more than the opposition in almost all categories.} 
    \label{fig:bias_in_appearance}
    \vspace{-\baselineskip}
\end{figure}
In debates critical of the ruling party, the hashtags tend to be issue-centric rather than party-centric. For example, hashtags like \#WillYogiSackMLA, and \#YogiWakeUp focus on individual incidents or politicians and don't necessarily indict the ruling party as a whole.
On the contrary, hashtags targeting the Opposition often portray them as either against the country or as disorganized and ineffective. Examples include \#CongInsultsDemocracy and \#RahulMocksForces, where the use of `Cong' (an abbreviation for the main opposition party) implies that the entire party is undermining democratic values or the armed forces. Further, hashtags like \#MamataLosesGrip or \#MayaDumpsCong indicate that the opposition parties are fractious and unreliable.
The full list of hashtags used in our analysis is shown in Appendix \ref{sec:additionaldetails}.

By analyzing the affiliations of panelists, whose names were extracted from the transcripts, we observe a discernible bias in the selection process for the show's panelists. As illustrated in Figure \ref{fig:bias_in_appearance}, there is a disproportionate tendency to invite spokespeople or supporters of the ruling party across various categories.

\subsection{Gender Bias}

\begin{figure}
    \centering
    \includegraphics[width=0.9\linewidth, clip=true, trim=0 25 0 10]{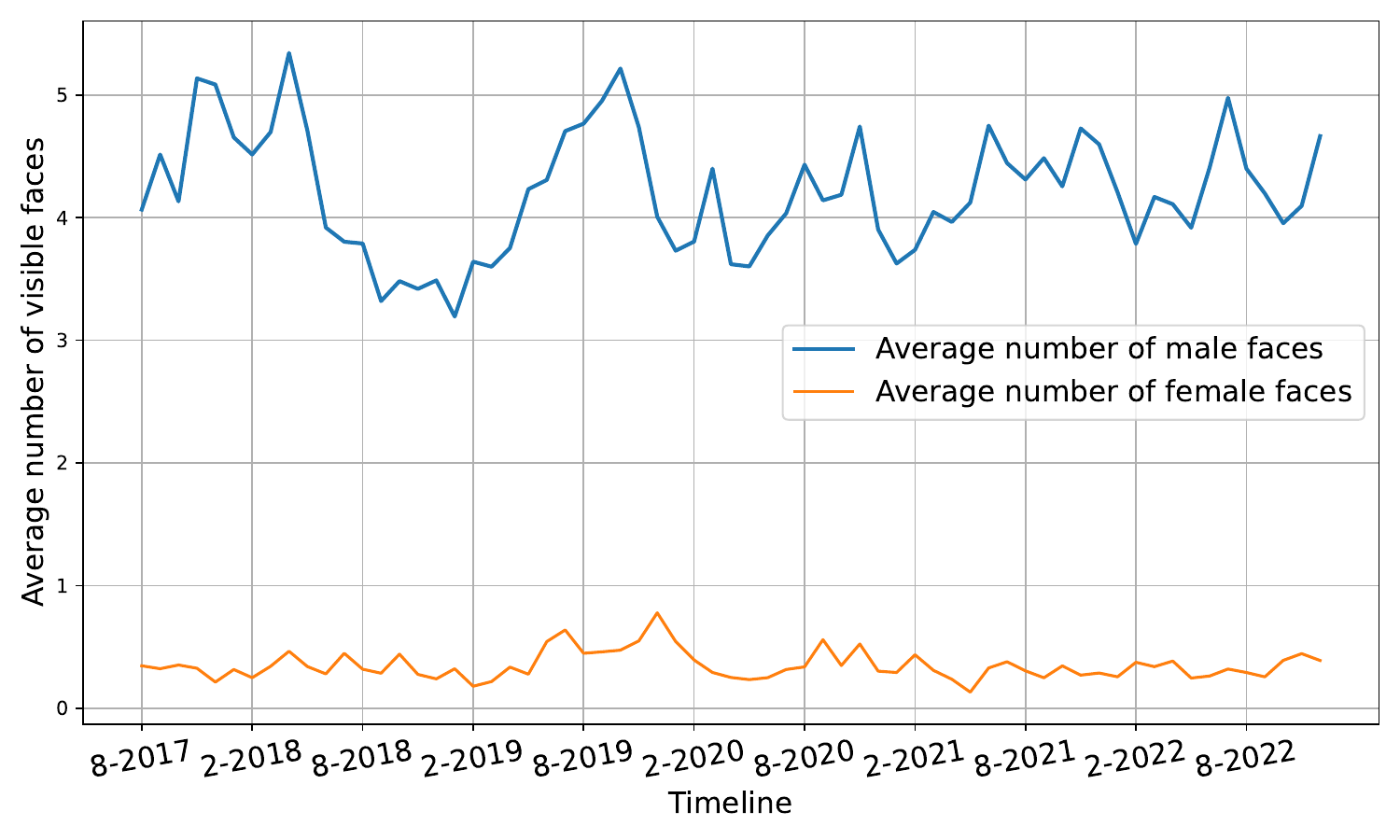}
    \caption{Average number of faces observed when a frame is randomly sampled from a video in the given month. Female guests are consistently underrepresented compared to their male counterparts.}
    \label{fig:female_lesser_males}
    \vspace{-\baselineskip}
\end{figure}

Figure \ref{fig:female_lesser_males} provides a temporal analysis of the gender distribution of faces visible during the debate videos, spanning a period of six years. The data unambiguously shows that females are consistently underrepresented when compared to their male counterparts. This trend is not isolated to specific periods but is persistent across the entire dataset's history.

We further delved into the issue by examining the representation of females in debates across various categories. Figures \ref{fig:more-female-faces} and \ref{fig:less-female-faces} highlight the top 5 and bottom 5 categories in terms of female representation, respectively. The data corroborates the presence of systemic gender bias. Notably, there are no categories where females constitute the majority. Although Bollywood-related debates are an outlier, having nearly 40\% of the panelists as women, in other categories, female presence is alarmingly sparse. For instance, in critical and often polarizing topics like the Citizenship Amendment Act (CAA) or the Kashmir issue, women make up only about 20\% of the panelists. This under representation becomes even more stark in debates about the Pulwama terror attack, where women occupy a mere 5\% of the screentime.

In addition to presence, we assessed the screen space allocated to each gender by measuring the average size of visible faces in square pixels. Our findings show that, on average, male faces occupy $3,798.51$ sq pixels, while female faces are allotted only $2,424.87$ sq pixels. This discrepancy is not an isolated occurrence but a consistent pattern over time, as illustrated in Appendix \ref{sec:additionaldetails}. The limited screen space for women, even when they are present, underscores the bias.

Our comprehensive dataset of 3,000 videos reveals that women account for a mere 7.5\% of the total screen time, which diminishes to 7.2\% in political debates. This underrepresentation is stark when compared to the presence of women in Indian politics, where females make up 14.32\% of Parliament members, and around 25\% of the internet population in India.

As we will discuss in Section \ref{sec:incivility}, there is a correlation between categories with lower female representation and higher levels of incivility. This correlation raises concerns about the quality of discourse and suggests that the gender imbalance may contribute to a more hostile debate environment. It also challenges the inclusivity of media channels in reflecting diverse viewpoints, especially on matters of national and societal significance.

\section{Incivility in the Debates}
\label{sec:incivility}

Indian television debates, particularly the one under study, are often marked by high levels of incivility and excessive dramatization, characteristics that can both entertain and polarize the audience. While these traits contribute to the show's popularity, they raise serious questions about the quality of public discourse and democratic debate in the country. In this section, we aim to quantify these elements of incivility using three carefully chosen metrics: \textit{(1) speech overlap, (2) use of foul language, and (3) instances of shouting}. \textit{Speech overlap} acts as a proxy for conversational decorum, with excessive overlap often indicative of a lack of respect for differing opinions. The use of \textit{foul language}, operationalized through detecting hateful language using Google's Perspective API~\cite{lees2022new}, directly reflects the tone and content of the debate, revealing any underlying animosities or prejudices. Lastly, the \textit{frequency of shouting} by the panelists offers insights into the emotional intensity of the debate, potentially correlating with heightened levels of aggression or antagonism. Collectively, these metrics provide a comprehensive lens to quantify and understand incivility in the complex setting of Indian TV debates.

\subsection{Overlapping Speech and Toxicity} \label{sec:methods_incivility}

The debates often elicit an emotional response from the panelists which either results in (1) panelists speaking over each other or (2) using foul speech to attack others' opinions~\cite{indiatomorrowDirtyGame}. 

\begin{figure*}
\centering    
    \begin{subfigure}[b]{0.33\textwidth}
        \centering
        \includegraphics[width=0.9\linewidth, clip=true, trim=0 20 0 10]{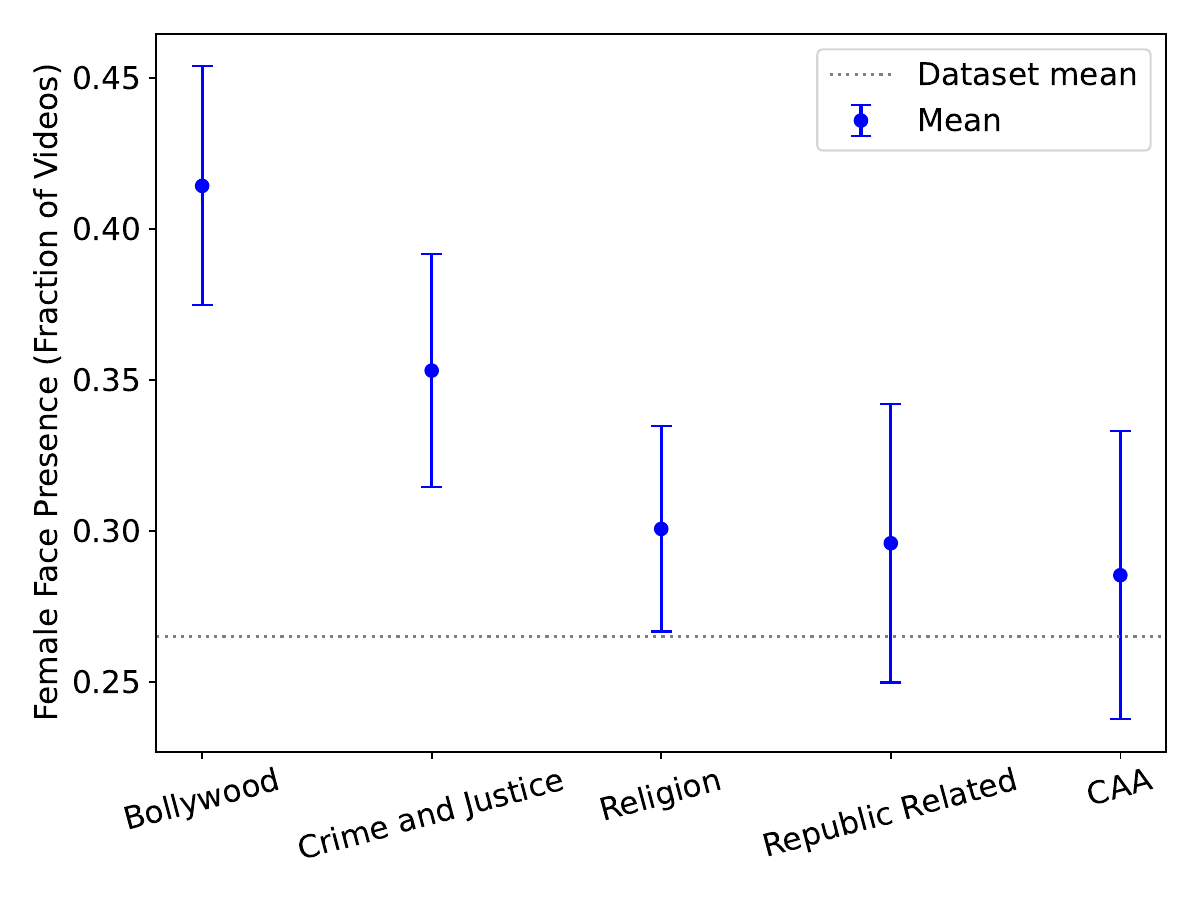}
        \caption{}
        \label{fig:more-female-faces}
        \vspace{-\baselineskip}
    \end{subfigure}
    \begin{subfigure}[b]{0.33\textwidth}
        \centering
        \includegraphics[width=0.9\linewidth, clip=true, trim=0 20 0 10]{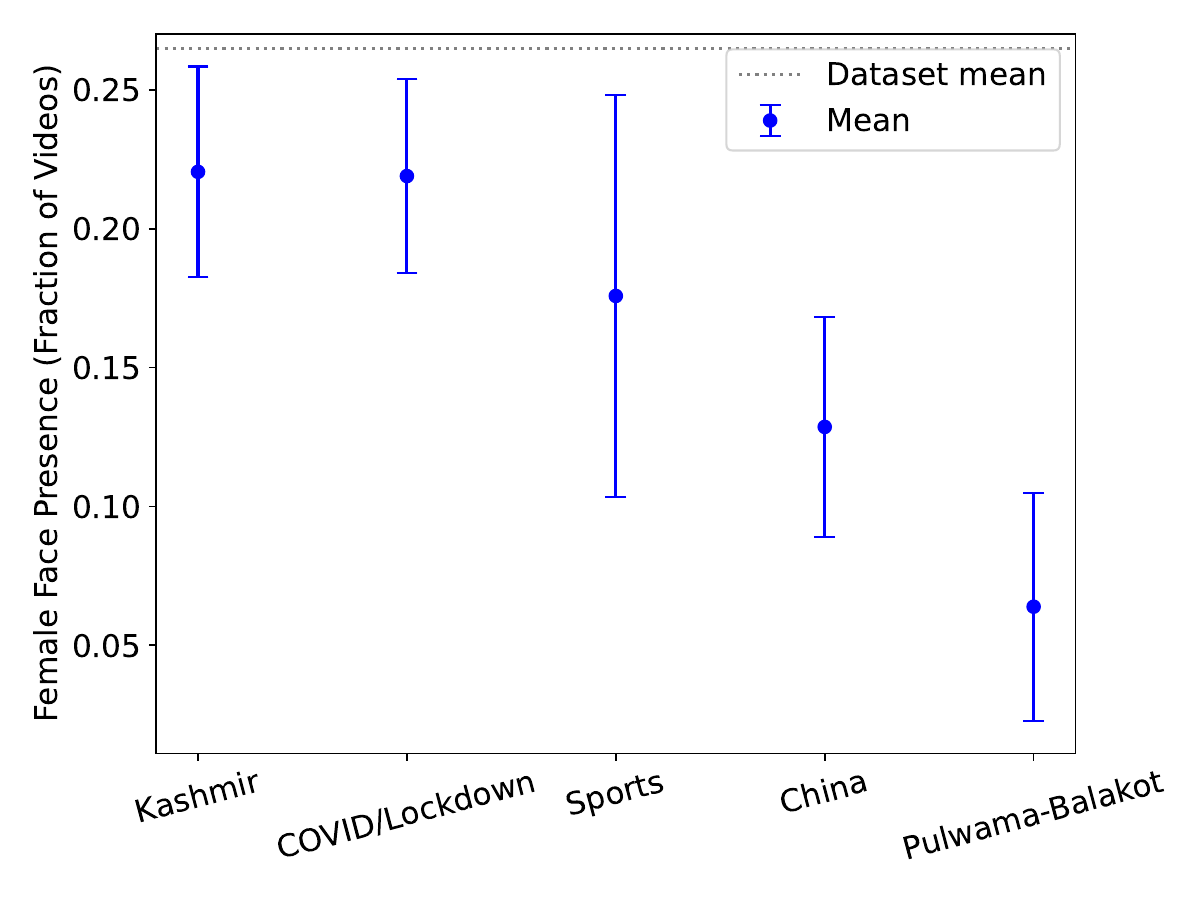}
        \caption{}
        \label{fig:less-female-faces}
        \vspace{-\baselineskip}
    \end{subfigure}
    \begin{subfigure}[b]{0.33\textwidth}
      \centering
      \includegraphics[width=0.9\linewidth, clip=true, trim=0 20 0 10]{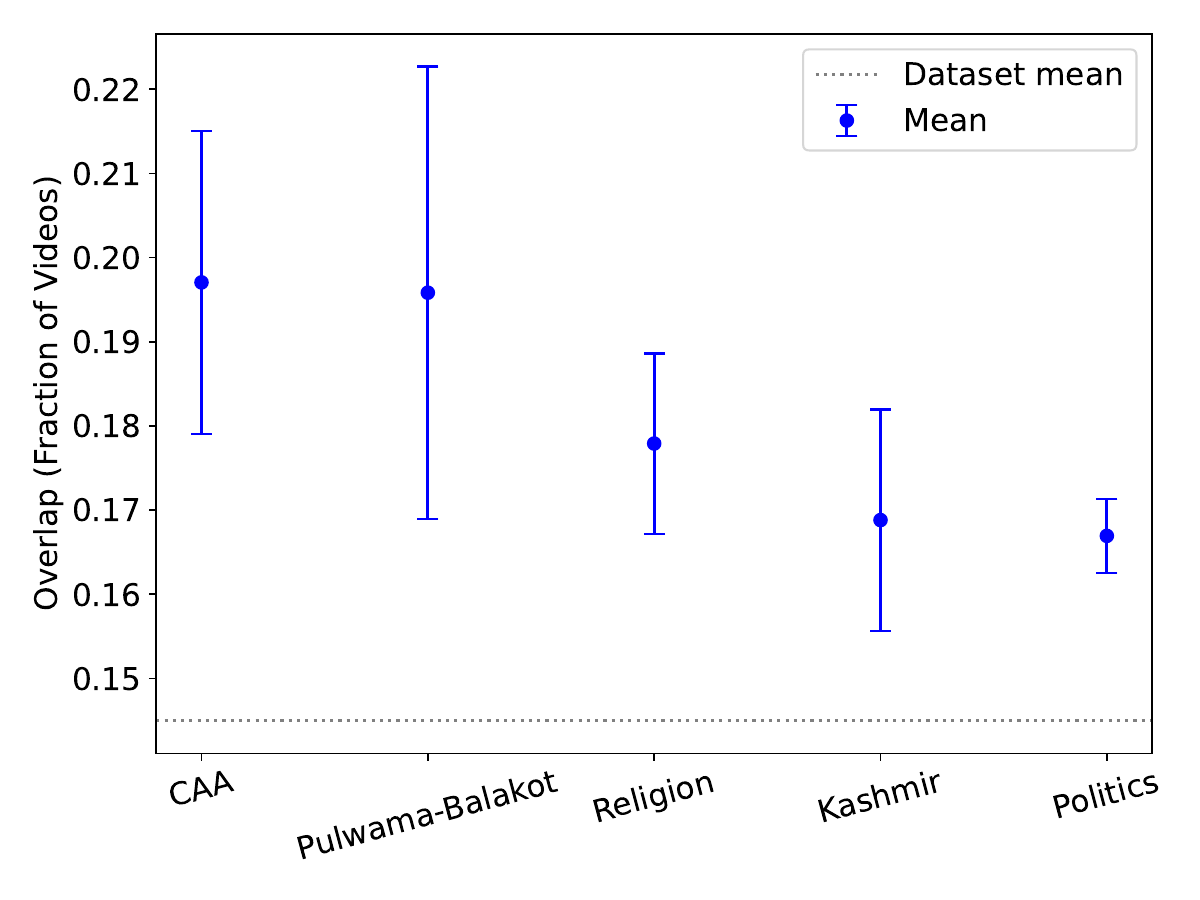}
      \caption{}
      \label{fig:incivil_frac_overlap_top}
      \vspace{-\baselineskip}
    \end{subfigure}
    \begin{subfigure}[b]{0.33\textwidth}
      \centering
      \includegraphics[width=0.9\linewidth, clip=true, trim=0 20 0 10]{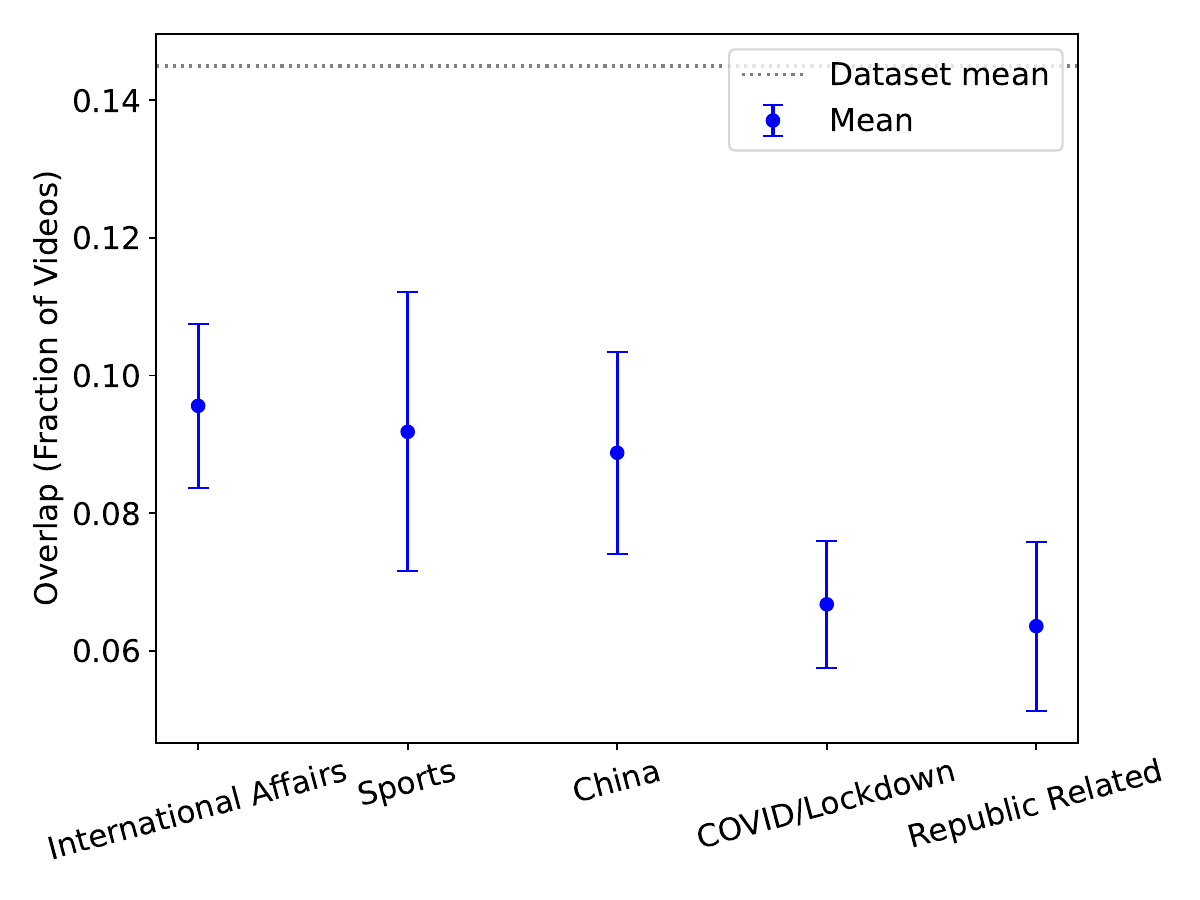}
      \caption{}
      \label{fig:incivil_frac_overlap_bottom}
      \vspace{-\baselineskip}
    \end{subfigure}
    \hfill
    \begin{subfigure}[b]{0.33\textwidth}
        \centering
        \includegraphics[width=0.9\linewidth, clip=true, trim=0 20 0 10]{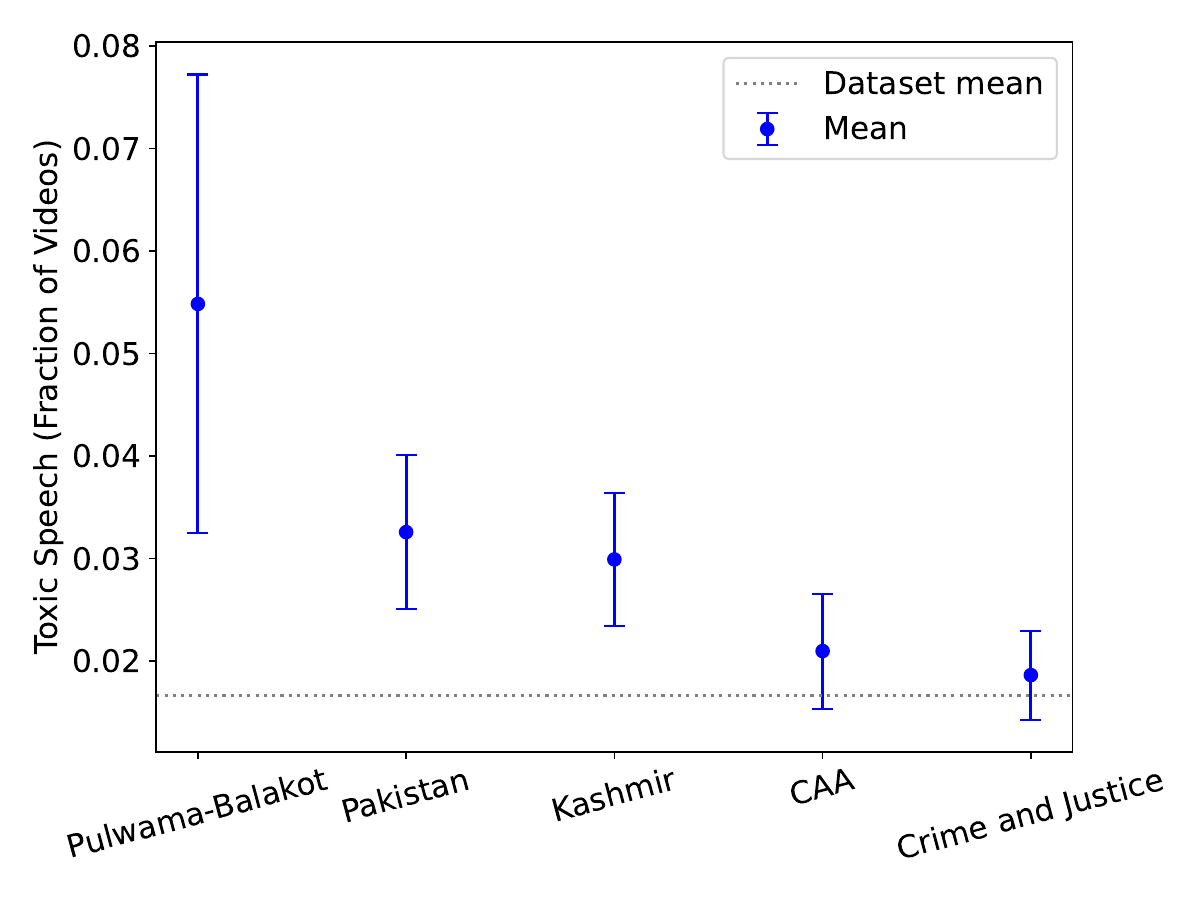}
        \caption{}
        \label{fig:incivil_frac_foul}
        \vspace{-\baselineskip}
    \end{subfigure}
    \hfill
    \begin{subfigure}[b]{0.33\textwidth}
        \centering
        \includegraphics[width=0.9\linewidth, clip=true, trim=0 20 0 10]{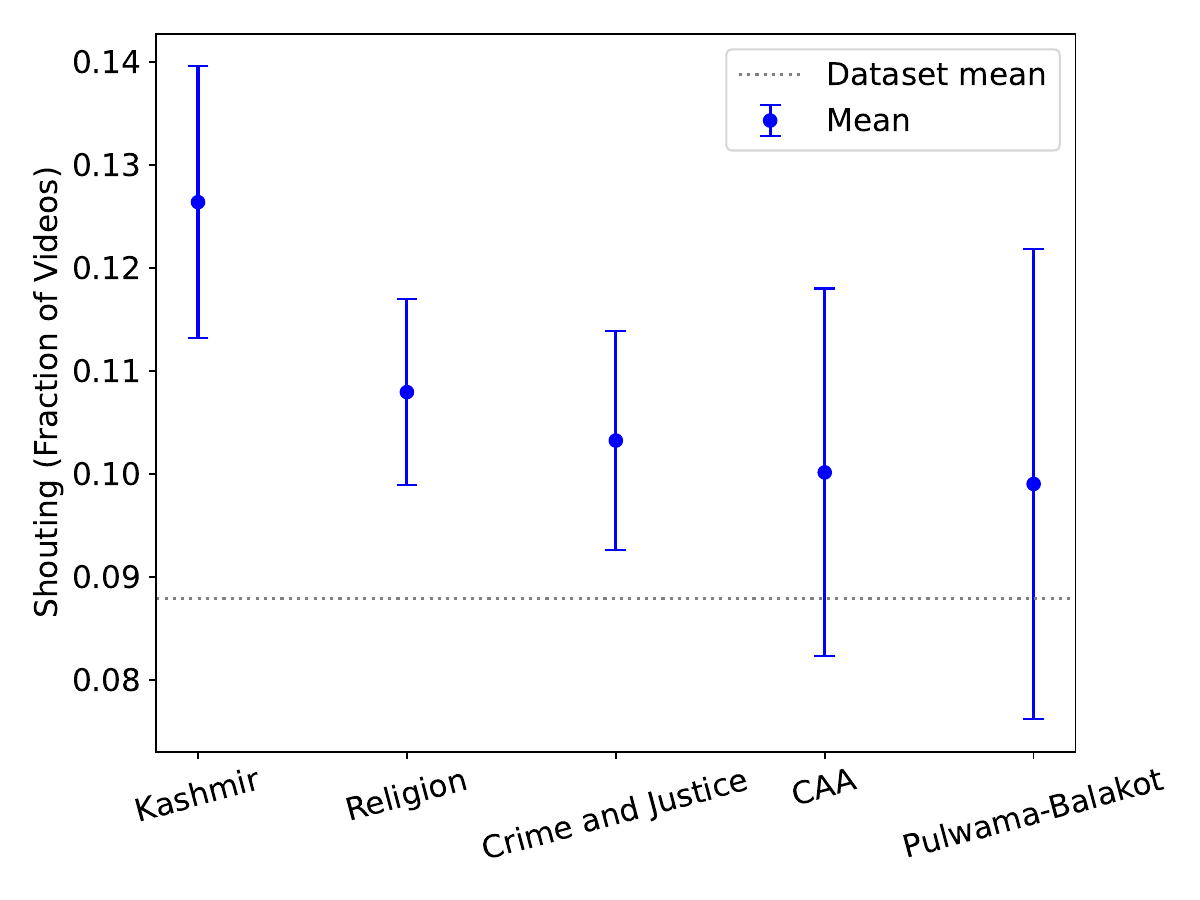}
        \caption{}
        \label{fig:shouting_fraction}
        \vspace{-\baselineskip}
    \end{subfigure}
    \caption{Confidence Intervals. (a) Top-5 categories with more females than average. (b) Bottom-5 categories with less females than average. (c) Fraction of the total duration of videos exhibiting overlapped speech for the top-5 categories, significantly exceeding the dataset's mean. The highest-ranking category has 20\% of video duration overlapping speech. (d) Fraction of the total duration of videos with overlapping speech for the bottom-5 categories, significantly below the dataset's mean. (e) Fraction of the total duration of videos with toxic speech in the top-5 most toxic categories. (f) Fraction of the total duration of videos with most shouting in the top-5 categories.}
    \label{fig:all_confidence_interval}
\end{figure*}

To identify \textbf{overlapping speech}, we follow the procedure outlined in Section \ref{sec:transcription_diarization}. 
Figures \ref{fig:incivil_frac_overlap_top} and \ref{fig:incivil_frac_overlap_bottom} show the top and bottom 5 categories which are significantly over or under the mean respectively.
They indicate a pronounced pattern of overlap in specific categories of debates, with particularly elevated levels observed in discussions revolving around contentious issues like the Citizenship Amendment Act (CAA), Kashmir, Politics, and Pulwama-Balakot events~\cite{siyech2019pulwama}, as well as Religion.
It is striking to note that in debates on the Pulwama terror attack, the CAA, and Kashmir, over 20\% of the discourse features overlapping speech. This suggests that these highly contentious issues are divisive and incite a breakdown in conversational decorum.
Conversely, we find markedly lower levels of incivility in debates related to International Affairs, COVID-19, the TRP Scam related to \channel, Sports, and Bollywood.
\begin{figure}
\centering    
    \begin{subfigure}[b]{0.49\textwidth}
      \centering
      \includegraphics[width=1\linewidth]{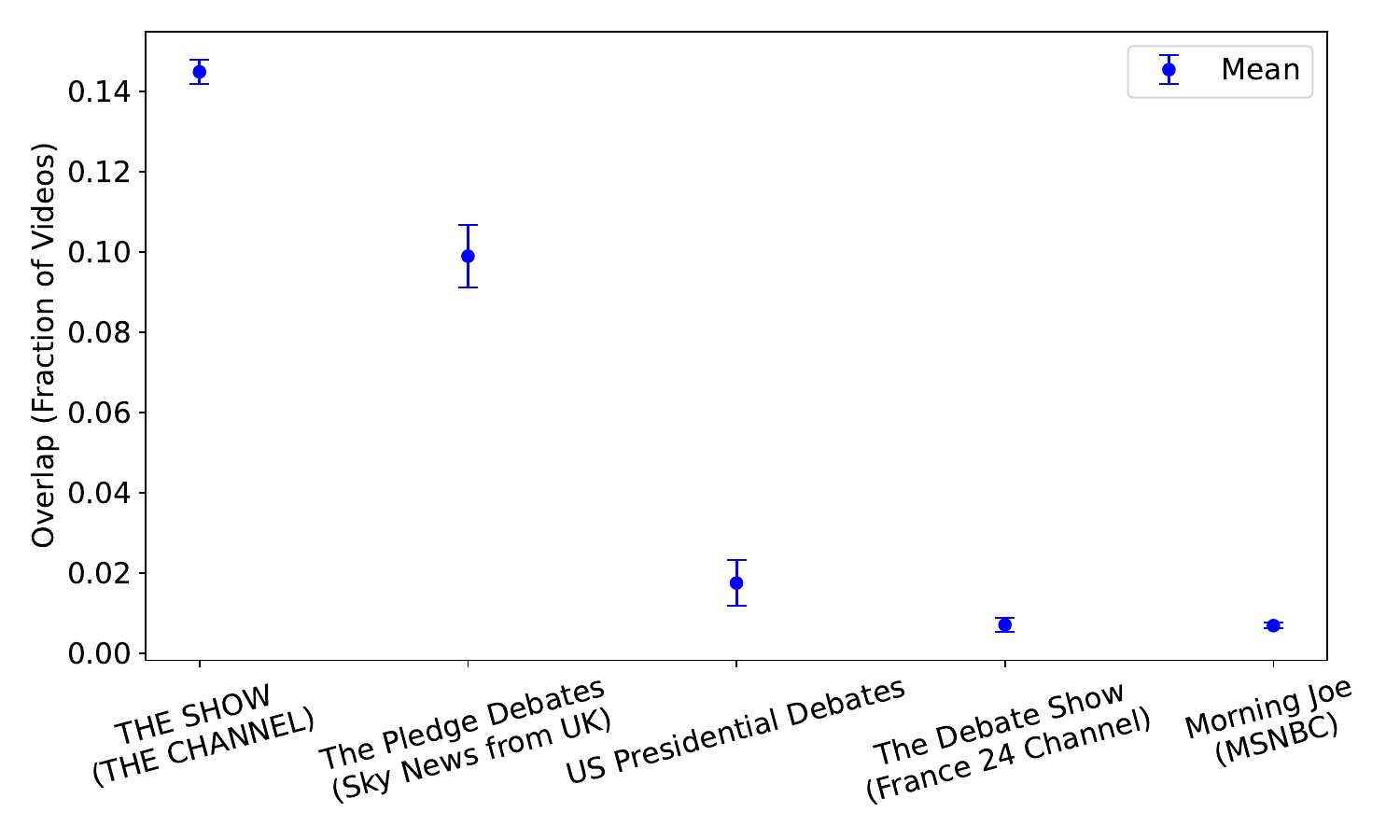}
      \caption{}
      \label{fig:overlap_other_debates}
    \end{subfigure}
    \hfill
    \begin{subfigure}[b]{0.49\textwidth}
          \centering
          \includegraphics[width=1\linewidth]{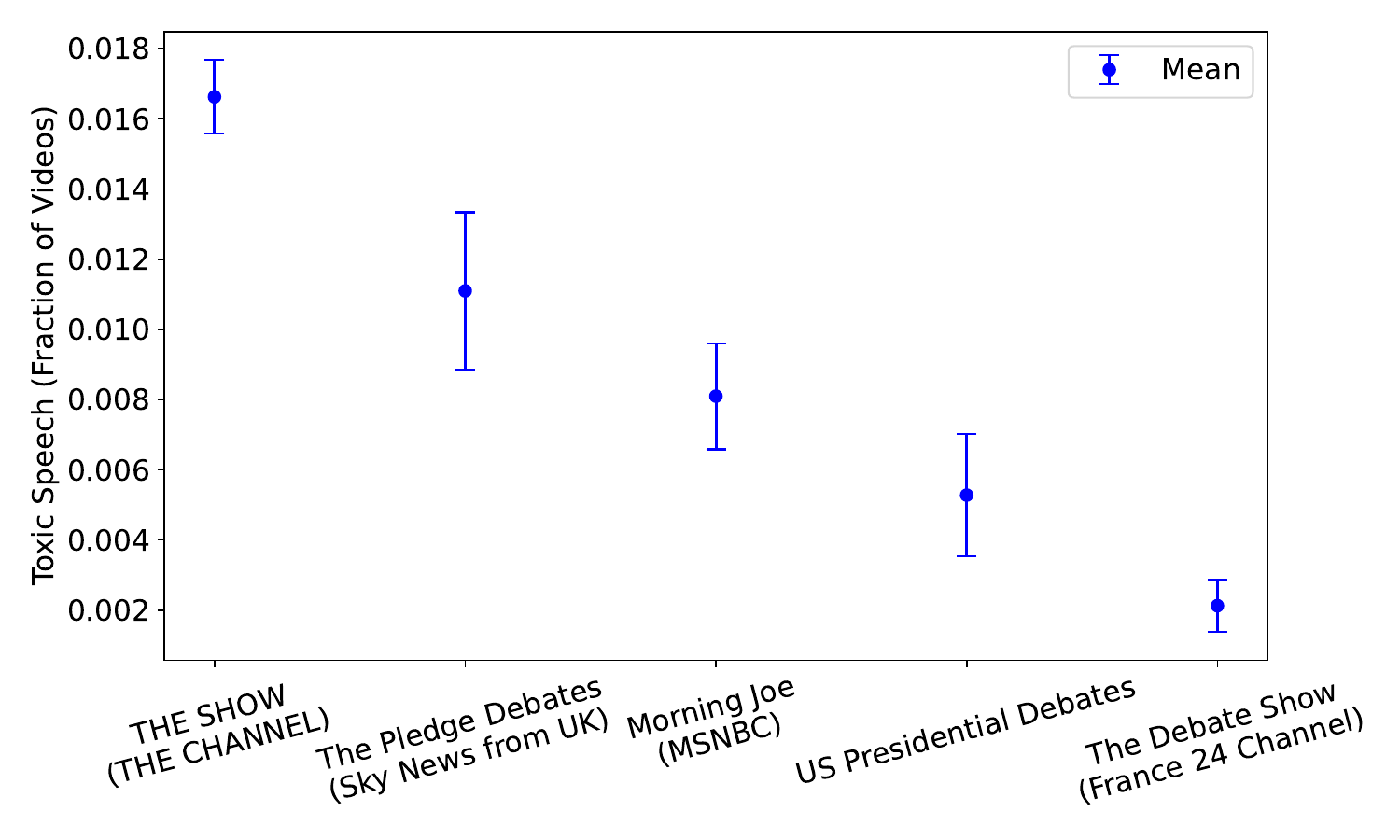}
          \caption{}
          \label{fig:toxic_other_debates}
    \end{subfigure}
    \hfill
    \caption{Comparison with other TV debate channels: (a) Fraction of video duration with overlapping speech. (b) Fraction of video duration with toxic speech.}
    \label{fig:comparing_other_debates}
\end{figure} 
We next turn our attention to the prevalence of toxic speech, specifically the use of \textbf{foul language}, in prime-time news debates. Contrary to what one might expect from a mainstream platform, the presence of toxic speech is not an aberration but rather an unsettling norm. To quantitatively measure toxicity, we employ the Perspective API~\cite{lees2022new}, which assesses text across multiple dimensions including toxicity, identity attack, insult, profanity, severe toxicity, and threat. Our analysis, detailed in Figure \ref{fig:incivil_frac_foul}, shows that an average of over 1\% of the duration across videos in our dataset contain some form of foul language. While this percentage may seem relatively low, it gains significance when considering the show's mass viewership, often in the millions. Most strikingly, the categories registering the highest toxicity levels are those discussing sensitive topics like Pakistan, Kashmir and terrorist attacks. These topics require the most thoughtful and nuanced discussion, yet they have been reduced to shouting matches and verbal attacks.

Elevated levels of incivility (captured through overlap speech and toxic speech) are not just isolated events but indicative of a broader trend that compromises the quality of public discourse. When panelists choose disruption over dialogue, they contribute to a media environment where aggressive and confrontational behaviour becomes the norm rather than the exception. 

\noindent\textbf{Generalizability:} \textit{Though the current study focuses on Indian TV debates, our pipeline is adaptable to other content on the web, specifically to debate shows in English.} To demonstrate its generalizability and establish baselines, we applied our pipeline to four English debate/panel-based shows: The Debate Show (France 24), The Pledge Debates (Sky News, UK), Morning Joe (MSNBC, US), and US Presidential Debates (2008-2020). Our analysis (Figure \ref{fig:comparing_other_debates}) compared overlapping speech and toxicity in these shows and found that the shows on \channel~have a statistically significantly higher incivility ($p < 0.01$) than all these shows. Refer to Appendix \ref{sec:other-dataset-performance} for the statistics and details on data collection for other debates.

\subsection{Detecting Shouted Speech}
To capture incivility holistically, it is imperative to not just study what is said but how it was said.
Shouting is another form of incivility used to overpower others' opinions in a debate.
Shouting detection in human speech is an established area of research ~\cite{pohjalainen2013detection}.

The Indian Broadcast News Debate (IBND) corpus~\cite{baghel2021automatic} contains news debates from \channel~ with annotations for shouted vs. normal speech. We used only the data corresponding to debates held on \channel~ since all our inferences will be performed on samples from the same domain.
Using the raw audio from videos in our dataset, we extract 26 MFCCs\footnote{Mel Frequency Cepstral Coefficients (MFCCs) of a signal are features which concisely describe the overall shape of an audio spectral wave.} per frame per audio file, with a frame size of 25ms and a gap of 10ms. On a per-audio level, we perform standard-scaling of these features and group frames into 1 second blocks. Inferences for shouting detection are performed on a per-second level.

We use a Convolutional Neural Network (CNN) to perform inference on per-second samples. The CNN  consists of four blocks. Each block contains a convolutional layer with a ReLU activation function, a max pooling layer for down-sampling, and a dropout layer for regularization and ends with a fully connected layer with a sigmoid activation function for binary classification. The CNN was compiled with the Adam optimization algorithm and binary cross-entropy as the loss function.
We tested our approach on the IBND dataset with an 80/20 train-test split, ensuring no data leakage by dividing on a per-audio basis. The model achieved 85\% accuracy and, with a high precision of 0.862, was deemed reliable for broader application. A majority voting system for continuous shouting further minimized false positives. The lower recall of 0.71 suggests that shouting instances may be underreported. Manual checks of randomly sampled shouting instances found no false positives. For validation of the classifier's performance, see Appendix \ref{subsec:shouted-speech-validation}.

Figure \ref{fig:shouting_fraction} shows the average percentage of time shouting occurs in each video, focusing on the top five categories. The complete plot for all categories is included in \supplement. Shouting occupies 9\% of the video duration on average, suggesting a notable departure from civil discourse. Categories like Kashmir, Religion, and Crime \& Justice are especially prone to high levels of shouting, corroborating the findings in Figures \ref{fig:incivil_frac_overlap_top}, \ref{fig:incivil_frac_overlap_bottom}, and \ref{fig:incivil_frac_foul}. This level of shouting, particularly in sensitive topics, underscores the emotionally charged nature of these debates. It raises questions about the efficacy of such discourse in fostering meaningful dialogue.

Additional analysis on panelist participation in shouting and incivility is detailed in the Appendix (Sections \ref{sec:appendix_shouting1}, \ref{sec:appendix_shouting2}). These results further corroborate the extensive presence of incivility and its correlation with debate dynamics.
\section{Discussion}

Our research employs a comprehensive toolkit, integrating state-of-the-art \textit{open-source tools} in computer vision, speech processing, and NLP, to analyze large quantities of video content. We apply this toolkit to a case study involving one of India's most-watched prime-time television debate shows, which garners over five million daily viewers. 
The show has received critique for its emphasis on strong nationalistic sentiments and its approach towards minority communities. By making our code public, we aim to encourage further research and analysis in diverse contexts. 

Our analysis uncovers significant bias and incivility within the debates, including a notable underrepresentation of women and a bias towards the ruling party. While there has been anecdotal evidence suggesting such biases, our research quantifies these biases. The act of delegitimizing opposition voices has far-reaching implications for the democratic discourse. Our analysis suggests that the use of sensationalism and dramatization may be a deliberate tactic rather than merely a byproduct of the show's popularity. Around 10\% of the debate time involves shouting, highlighting an environment that is antithetical to civil discourse. 

Television's significant influence on public opinion is concerning when coupled with the biases we've identified~\cite{blumler1970political}. This becomes even more alarming considering that opposition coalitions have started boycotting certain television hosts based on similar criticisms~\cite{indiatodayINDIABloc}, potentially furthering polarization. When millions rely on such a low-quality platform for political insights, the spread of biased information undermines democratic processes and could lead to a misinformed electorate. The high ratings of such shows despite their evident flaws introduce a complex paradox. It challenges the simplistic notion that the media merely reflects public opinion, suggesting that it may play a role in shaping/distorting it. 

Overall, our findings offer more than an academic contribution; they signal an urgent call to action. They serve as a critical resource for researchers studying media ethics, democratic governance, and societal polarization. Our work raises complex questions about the ethical responsibilities of media in a democracy and the influence of media on public opinion. These issues warrant investigation and should be of concern to policymakers, civil society organizations, and the public at large.

\noindent\textbf{Limitations}: \textit{(i) Scope}: Our study is limited to a single prime-time news debate show and may not apply to more informal content like TikTok videos, which have highly variable discourse quality and nature. (ii) \textit{Manual Annotation}: The need for manual annotation in categorizing videos and identifying panelists limits scalability and could introduce bias. (iii) \textit{Technical Constraints}: Our work is constrained by the accuracy and potential biases of the classifiers, with the risk of compounded errors throughout the pipeline stages.

\noindent\textbf{Ethics Statement}: While our toolkit makes large video datasets more tractable for analysis, the potential for misuse is present; for example, the ability to index and search entire video archives could pose significant privacy risks. As with any tool, the ethical implications of its application should be carefully considered according to the use case. Considering that politicians and political analysts are public figures, and taking into account the significance of research in comprehending the language employed in political debates and its consequences, we believe our work conforms to acceptable standards of privacy \cite{doherty_ethics}.

\noindent\textbf{Future Work}.
This study merely scratches the surface of what can be achieved with automated, large-scale analysis of televised debates. We have not fully utilized diarization data due to clustering challenges. While speech embeddings have been tested, they need refinement for practical use. Future work could use diarization for deeper analyses like anchor bias or systemic media bias Overall, while our study has limitations, it offers a pioneering approach to multimedia content analysis, setting the stage for more comprehensive, automated methods in the future.

\begin{acks}
Hitkul is supported by TCS Research Scholar Program.
Kiran Garimella's research is funded by grants from the National Science Foundation, Knight Foundation and Google.
\end{acks}
\bibliographystyle{ACM-Reference-Format}
\balance
\bibliography{kdd_references}



\appendix

\section{Performance on other datasets}
\label{sec:other-dataset-performance}
To demonstrate the generalizability of our pipeline, we applied it to four additional English-language debate series:

\begin{itemize}
  \item \href{https://www.youtube.com/playlist?list=PLCUKIeZnrIUlLhXw4GoHFlUpFidIosXAT}{The Debate Show} (hosted on the France 24 Channel): Analyzed $216$ videos from their YouTube playlist.
  \item \href{https://www.youtube.com/@thepledge/videos}{The Pledge Debates} (hosted on Sky News in the UK): Focused on full-length debate videos from their YouTube channel, specifically selecting those exceeding 20 minutes in duration.
  \item \href{https://www.youtube.com/playlist?list=PLDIVi-vBsOEwnLoiImPnGhZfKvXzTlbee}{Morning Joe} (hosted on MSNBC in the US): Curated a set of videos from their YouTube playlist, including only those longer than 30 minutes to concentrate on complete episodes of the main show, resulting in $403$ videos for our study.
  \item US Presidential Debates (from 2008-2020): Compiled $38$ debate videos, encompassing the main presidential and vice-presidential debates from 2008-2012, as well as intra-party candidate-nomination debates.
\end{itemize}

We conducted a two-tailed t-test with a 95\% confidence interval to compare the overlap speech in debates from \gitdiff{Republic TV}{\channel} with the other debates mentioned. We found that the overlap speech in \gitdiff{Republic TV debates}{\tvshow} is statistically greater than in all the other TV debates listed above. Detailed results can be found in Table~\ref{tab:hypo_overlap_other_debates}. Similarly, when analyzing toxicity levels, we determined that \gitdiff{Republic TV debates}{\tvshow} exhibits statistically higher toxicity compared to debates from France 24, the US Presidential Elections, and Morning Joe. For a comprehensive breakdown, refer to Table \ref{tab:hypo_toxic_other_debates}.

\begin{table*}[h]
    \caption{\textbf{One-Tailed \textit{t}-test to check difference between distribution of overlap speech and toxicity between \channel~ vs other shows is statistically significant, we report \textit{t}-stat for $\alpha=$ 0.05}}
    \label{tab:hypo_incivility_other_debates}
    \begin{subtable}{.5\linewidth}
      \centering

      \caption{Two-Tailed t-test for Overlap Speech}
      \label{tab:hypo_overlap_other_debates}
     {\small
\begin{tabular}{@{}p{2.8cm}@{\hspace{5pt}}ccccc@{}}
    \toprule
    \textbf{Debate} & \textbf{$Mean$} & \textbf{\textit{t}-stat} & \textbf{\textit{p}-value}\\
    \midrule
    	\channel & 0.1448 & NA & NA \\
	France 24 & 0.0076 & 23.2468 & 5.98-110 \\
	Sky News UK & 0.0984 & 5.1663 & 2.55-07 \\
	US Presidential Elections & 0.0175 & 9.9175 & 8.21-23 \\
	Morning show with Joe & 0.0069 & 35.2401 & 4.41-230 \\
    \bottomrule
\end{tabular}
}
    \end{subtable}%
    \begin{subtable}{.5\linewidth}
      \centering

      \caption{Two-Tailed t-test for Toxicity}
      \label{tab:hypo_toxic_other_debates}
      {\small
\begin{tabular}{@{}p{2.8cm}@{\hspace{5pt}}ccccc@{}}
    \toprule
    \textbf{Debate} & \textbf{$Mean$} & \textbf{\textit{t}-stat} & \textbf{\textit{p}-value}\\
    \midrule
	\channel & 0.0166 & NA & NA \\
	France 24 & 0.0021 & 6.9221 & 5.43-12 \\
	Sky News UK & 0.0110 & 1.7640 & 0.0778 \\
	US Presidential Elections & 0.0053 & 2.4801 & 0.01 \\
	Morning show with Joe & 0.0081 & 5.9825 & 2.44-09 \\
    \bottomrule
\end{tabular}
}
    \end{subtable} 
\end{table*}

    

\begin{table}[htbp]
\small
\caption{Frequency of major and minor labels across the debate videos used in our analysis}
\label{tab:simple_stats}
\begin{tabular}{p{0.40\linewidth}p{0.20\linewidth}p{0.20\linewidth}}
\hline
Category & Videos where present as major label & Videos where present as minor label \\
\hline
Politics & 1209 & 739 \\
Religion & 216 & - \\
Crime and Justice & 190 & 262 \\
International Affairs & 181 & 128 \\
COVID/Lockdown & 181 & - \\
Pakistan & 155 & 47 \\
Bollywood & 140 & - \\
Kashmir & 134 & 3 \\
Political Scams & 128 & - \\
Citizenship Amendment Act & 87 & - \\
Republic TV related & 77 & - \\
Economy & 76 & 3 \\
China & 58 & 6 \\
Defense \& Terrorism & 50 & 288 \\
Farmers Protest issue & 42 & - \\
Pulwama-Balakot & 39 & - \\
Sports & 29 & - \\
Education & 8 & - \\
Anti-Opposition & - & 599 \\
State level politics & - & 548 \\
Supporting-BJP & - & 160 \\
SSR\_Case & - & 78 \\
Anti-BJP & - & 61 \\
Ram Mandir Babri Masjid & - & 59 \\
Russia-Ukraine & - & 49 \\
Triple Talaq & - & 15 \\
\hline
Total & 3000 &  \\
\hline
\end{tabular}
\end{table}


\section{Validation Experiments}
\label{subsec:shouted-speech-validation}
\textbf{Validation for classification of speech into shouted and non-shouted categories}: We manually annotated 50 audio samples from our dataset, classifying them as shouted or non-shouted. Our classifier achieved a precision of 0.91 and a recall of 0.75. The Indian Broadcast News Debate (IBND) dataset \cite{baghel2021automatic}, which includes debates from \channel~ with shout annotations, showed our classifier had a precision of 0.86 and a recall of 0.71 on 62,375 test samples. Given the IBND dataset shares our dataset's domain, these results suggest comparable performance on our data.

\section{Additional Analysis}
\subsection{Co-attendance Networks of Panelists}
\label{sec:network}

\begin{figure}[h]
        \centering
        \includegraphics[width=0.9\linewidth, clip=true, trim=0 25 0 20]{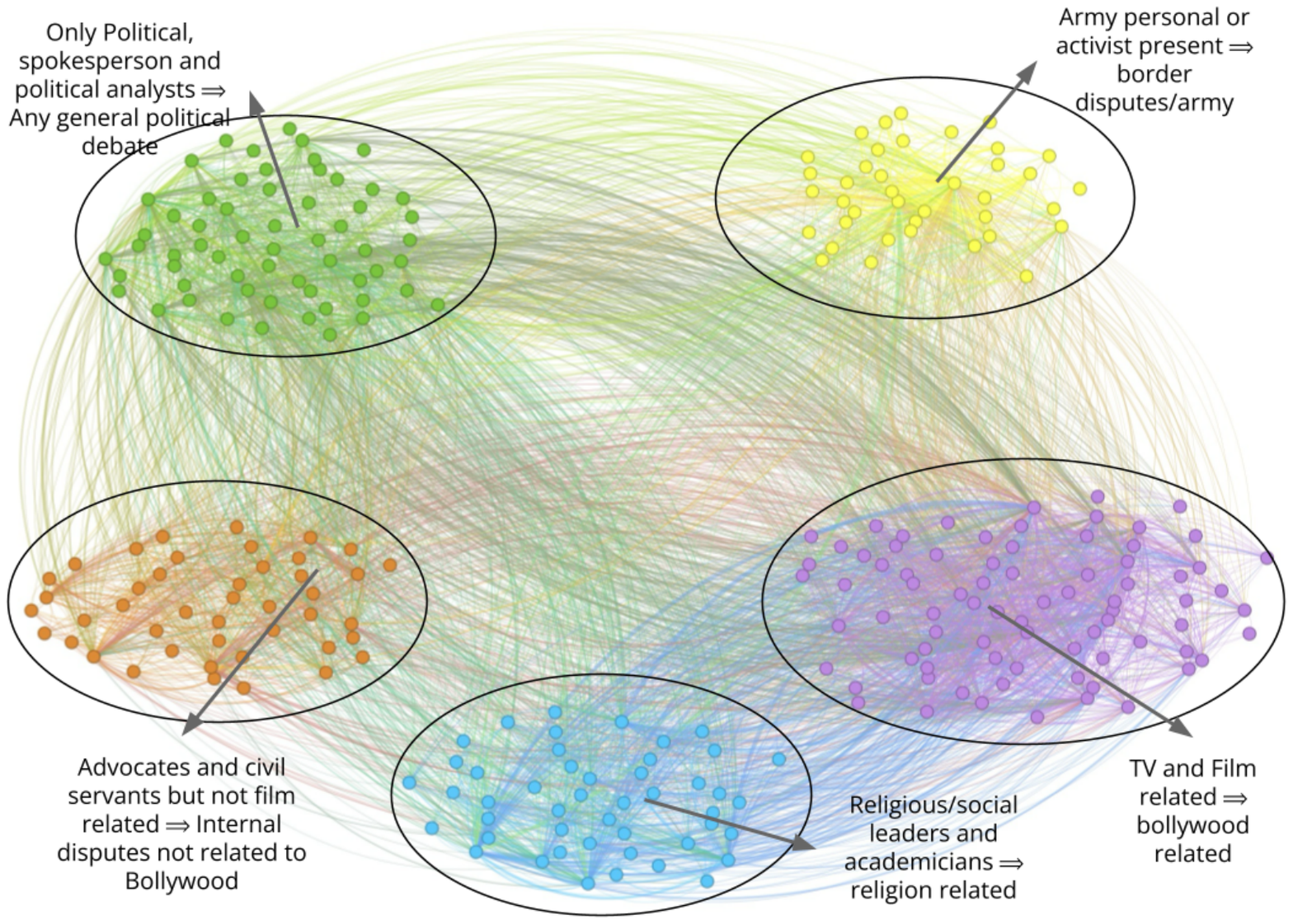}
        \caption{Panelist co-occurrence network}
        \label{fig:network_affiliation}
\end{figure}

Using panelists information we coded in Section \ref{sec:panelist_names}, we created a co-occurrence network between the panelists. If two panelists appeared together in a debate, they were connected by an edge.
We found that such a network (Figure \ref{fig:network_affiliation}) 
was clustered along categories and occupations of the panelists, indicating that the show invites specific panelists based on topics of discussion. The five communities were automatically identified using the Louvain method for community detection.  
(1) Orange: Found occupation like Advocate, civil servants but not film related occupation: Not related to Bollywood internal disputes 
(2) Blue: All religious/social leaders and academic people: Something related to religion 
(3) Pink: All TV and film related people: related to Bollywood
(4) Yellow: Army related personal, activists: related to border disputes/army
(5) Green: Only politician, spokesperson and analyst: Any general political debate

\subsection{Participants involved in shouting}
\label{sec:appendix_shouting2}

\begin{figure}[ht]
    \centering
    \includegraphics[width=\linewidth, clip=true, trim=0 20 0 15]{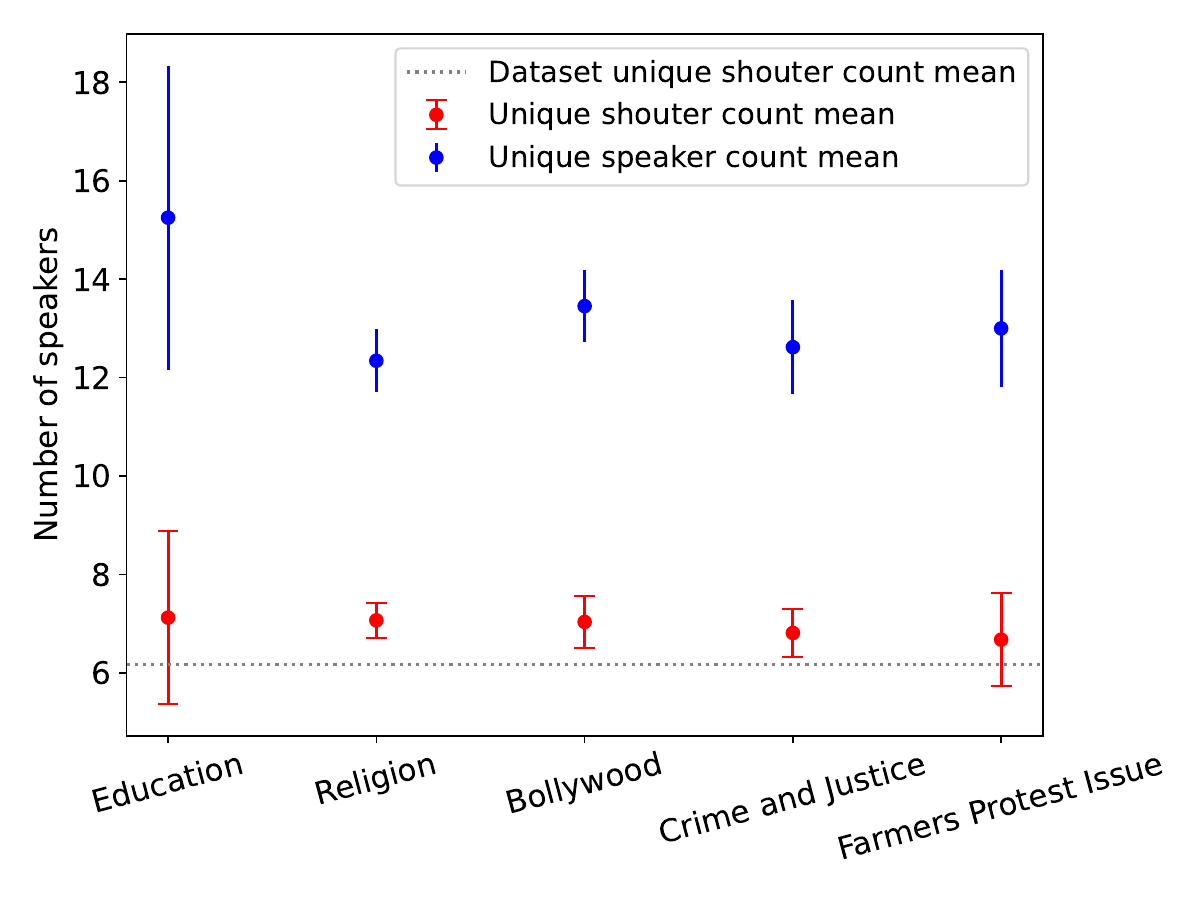}
    \caption{Average count of panelists engaged in shouting (in red) compared to the total panelist count (in blue) for top 5 categories with the highest incidence of shouting. The data indicates that 50\% of panelists in these categories participate in shouting behavior.}
    \label{fig:shouting_people}
\end{figure}

We look at the number of people participating in the shouting. By matching the shouting segments with the diarized text, we identify the speakers who shouted. We wanted to understand whether the debates are being derailed by a small group of people or many panelists have to engage in such behavior to have their voices heard. Figure \ref{fig:shouting_people} shows the top five categories ordered by the average number of panelists engaging in shouting along with the number of speakers on average in each category. We find that, roughly half of the participants engage in shouting.
It is also important to note that these categories with the highest number of shouting panelists are very different from the results we found in the rest of the figures documenting incivility (Figures \ref{fig:incivil_frac_overlap_top}, \ref{fig:incivil_frac_overlap_bottom},  \ref{fig:incivil_frac_foul}, and \ref{fig:shouting_fraction}).

\section{Additional Details}
\label{sec:additionaldetails}

\begin{figure}[h]
        \centering
        \includegraphics[width=1\linewidth, clip=true, trim=0 25 0 10]{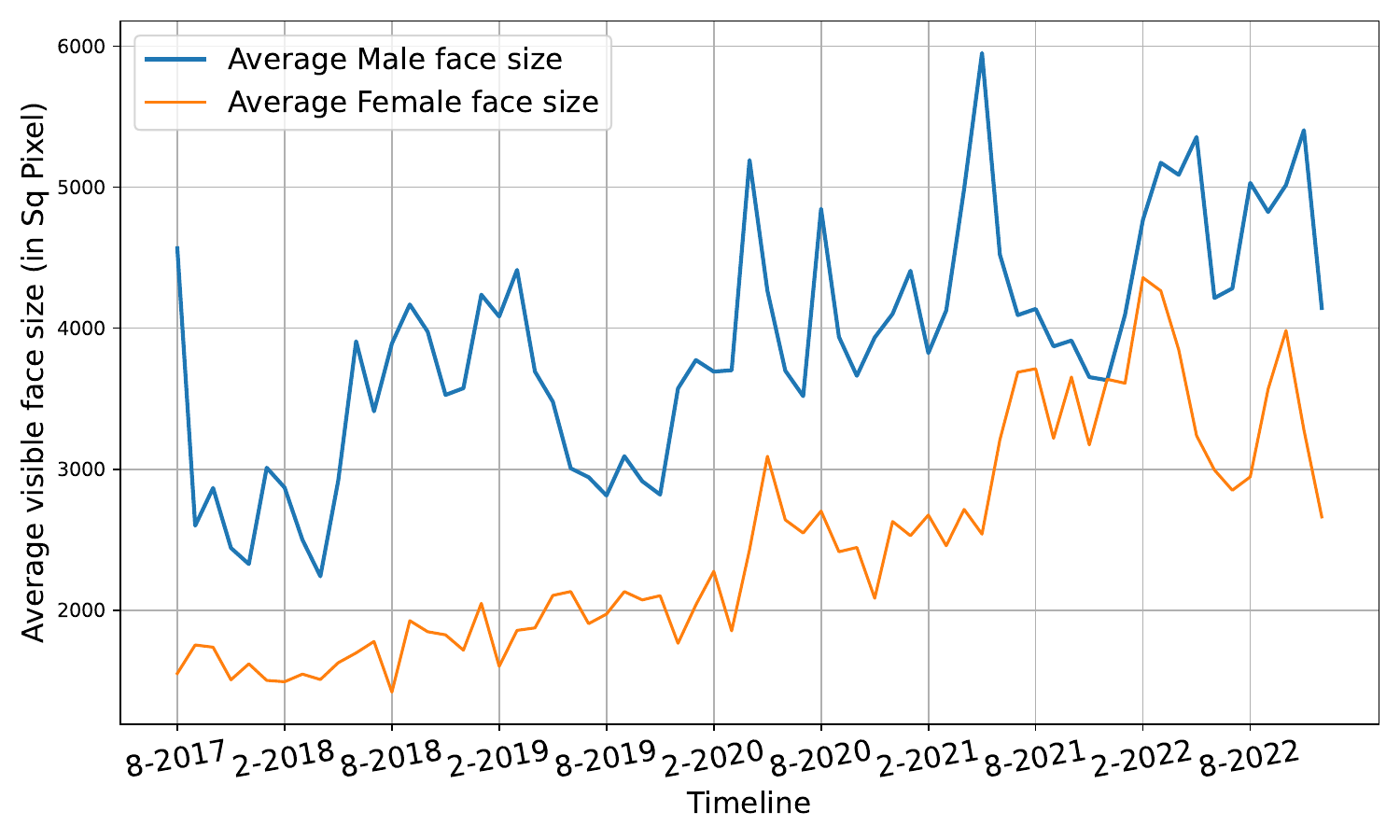}
        \caption{Average size of faces (Males: 3798 sq pixels, Females: 2424 sq pixels}
        \label{fig:female_size_lesser_males}
\end{figure}

\paragraph{Label Frequency:} Major and minor label distributions in our debates are in Table \ref{tab:simple_stats}.
\paragraph{Hashtag Analysis:} The compilation of hashtags employed in our bias-related analysis is provided in Table \ref{tab:hashtag_bias}.
\paragraph{Keyword Extraction:} The specific words utilized to extract sentences pertaining to the Opposition and Ruling parties for our bias analysis are listed in Table \ref{tab:classifier_dataset_keywords}.
\paragraph{Gender Disparity in Screen Space:} Gender-based screen space differences, favoring males, are shown in Figure \ref{fig:female_size_lesser_males}.

\paragraph{Method for quantifying foul speech:} We processed debate video transcripts, which consist of sequential utterances by different speakers. Using the Perspective API, we assessed each utterance for categories like toxicity, severe toxicity, profanity, insult, threat, or identity attack. Utterances with a probability over 0.5 in any category were marked as foul speech. 

\begin{table}[htbp]
    \caption{Hashtags showcasing the level of scrutiny between videos in Anti-Ruling-party vs Anti-Opposition videos}
    \label{tab:hashtag_bias}
    \begin{tabular}{|p{0.4\columnwidth}|p{0.5\columnwidth}|}
        \hline
        \textbf{Hashtags used in Anti-Ruling-party videos} & \textbf{Hashtags used in Anti-Opposition videos}\\
        \hline
        BaggaTweetArrest, YogiWakesUp, GovernorRightorWrong, WillYogiSackMLA, FightForAsifa, SadhviBackGodse, SackBJPBrat, RepublicVsBJPMLA, YogicopsStung, BJPWakeUpCall & CongRapeComment, MayaDumpsCong, CongVsCitizens, ConginsultsDemocracy, ECBansMamata, MamataLosesGrip, AAPForFreebies, KejriwalMinisterArrested, NeechPolitics, VadrasMustGo, RahulMocksForces, CongresslIsOver
\\
        \hline
    \end{tabular}

\end{table}
\begin{table}[htbp]
    \centering
    \caption{Keywords}
    \begin{tabular}{|p{0.45\columnwidth}|p{0.45\columnwidth}|}
        \hline
        \textbf{Ruling party Specific Words} & \textbf{Opposition Specific Words}\\
        \hline
        modi, narendra, shah, amit, yogi, adityanath, bjp & rahul, vadra, sonia, priyanka, gandhi, kejriwal, congress\\
        \hline
    \end{tabular}
    
    \label{tab:classifier_dataset_keywords}
\end{table}

\end{document}